\documentclass[structabstract]{aa}  

\usepackage{graphicx}
\usepackage{txfonts}

\usepackage{natbib}
\bibpunct{(}{)}{;}{a}{}{,} 

\begin{document}

   \title{The Galactic WC stars}

   \subtitle{Stellar parameters from spectral analyses indicate a new evolutionary sequence}

   \author{A. Sander
          \and
          W.-R. Hamann
          \and
          H. Todt
          }

   \institute{Institut f\"ur Physik und Astronomie, Universit\"at Potsdam,
              Karl-Liebknecht-Str. 24/25, D-14476 Potsdam, Germany\\
              \email{ansander@astro.physik.uni-potsdam.de, 
                     wrh@astro.physik.uni-potsdam.de}
             }

  \date{Received 5 August 2011; accepted 30 January, 2012}

\abstract{
  The life cycles of massive stars from the main sequence to their
explosion as supernovae or gamma ray bursts are not yet fully clear, and
the empirical results from spectral analyses are partly in conflict with
current evolutionary models. The spectral analysis of Wolf-Rayet stars
requires the detailed modeling of expanding stellar atmospheres in
non-LTE. The Galactic WN stars have been comprehensively analyzed with
such models of the latest stage of sophistication, while a similarly
comprehensive study of the Galactic WC sample remains undone.
}{
  We aim to establish the stellar parameters and mass-loss rates of the 
Galactic WC stars. These data provide the empirical basis of
studies of (i) the role of WC stars in the evolution of massive stars,
(ii) the wind-driving mechanisms, and (iii) the feedback of WC stars as
input to models of the chemical and dynamical evolution of galaxies.
}{
  We analyze the nearly complete sample of un-obscured Galactic WC stars,
using optical spectra as well as ultraviolet spectra when available. The
observations are fitted with theoretical spectra, using the Potsdam
Wolf-Rayet (PoWR) model atmosphere code. A large grid of line-blanked
models has been established for the range of WC subtypes WC4 -- WC8, and
smaller grids for the WC9 parameter domain. Both WO stars and WN/WC 
transit types are also analyzed using special models.
}{
  Stellar and atmospheric parameters are derived for more than 50
Galactic WC and two WO stars, covering almost the whole Galactic WC
population as far as the stars are single, and un-obscured in the
visual. In the Hertzsprung-Russell diagram, the WC stars reside between
the hydrogen and the helium zero-age main sequences, having luminosities
$L$ from $10^{4.9}$ to $10^{5.6}\,L_{\odot}$. The mass-loss
rates scale very tightly with $L^{0.8}$. The two WO stars
in our sample turn out to be outstandingly hot ($\approx$200\,kK) and 
do not fit into the WC scheme. 
}{
  By comparing the empirical WC positions in the Hertzsprung-Russell
diagram with evolutionary models, and from recent supernova statistics,
we conclude that WC stars have evolved from initial masses between 20
solar masses and 45\,$M_{\odot}$. In contrast to previous assumptions, 
it seems that WC stars in general do not descend from the most massive 
stars. Only the WO stars might stem from progenitors that have been initially
more massive than 45\,$M_{\odot}$.
}


\maketitle


\section{Introduction}
  \label{sec:intro}

  Wolf-Rayet (WR) stars represent advanced evolutionary stages in the
evolution of massive stars. Owing to their high temperatures and high 
mass-loss rates, WR stars provide a large amount of ionizing photons,
mechanical momentum, and matter to their environment, making WR stars a
keystone in the cosmic evolution. Because of the extreme non-local thermal
equilibrium (non-LTE) situations in the expanding atmospheres, their spectral
analysis only became possible after adequate, powerful and complex model 
atmosphere codes had been established. Since the early 90s, these tools have
been available. Early analyses have been published, e.g., by \citet{Hillier87},
\citet{SHW89}, \citet{HKW93}, and \citet{HKW95} for the WN stars, i.e.\ WR stars with
strong nitrogen emission. For the WC subclass, which are characterized by strong
carbon emission-lines, wind models have been applied by, e.g.,
\citet{Hi89}, \citet{HLKW92}, and \citet{KH92}.

  The models have since been significantly improved. With the inclusion of 
iron line blanketing and microclumping \citep{HiMi99,GKH02}, a new generation of 
models have allowed better spectral fits, and led to a revision of the empirical 
stellar parameters. A comprehensive re-analysis of the Galactic WN stars was 
performed by \citet{Adriane06}. Basically all WN stars in the Large Magellanic 
Cloud have been analyzed by \citet[][and in prep.]{Ute08}. \citet{DianaSMC}
studied the WR stars in the Small Magellanic Cloud. In contrast to the
WN spectra, the WC spectra turned out to be much harder to reproduce.
Our model spectra can fit most, but not all of the observed features,
leaving some ambiguity about the choice of the best-fit model.

  The present study is based on models that are not hydrodynamically
consistent. The mass-loss rate and wind velocity field are instead treated as 
free parameters. The only hydrodynamical WC model that has been constructed so far
is that developed to study the prototypical WC5 star WR\,111 \citep{GH05}. This 
model arrived at a much higher stellar temperature (140\,kK) and smaller mass-loss
rate than the semi-empirical approach of the present paper. Using this same model,
\citet{GKH02} had obtained $T_{*}$\,=\,85\,kK for WR\,111,
while only 62\,kK had been derived with non-blanketed models \citep{KH95}.

  In the afore mentioned paper of \citet{KH95}, 25 Galactic WC stars of
subtypes WC4 to WC8 were analyzed with models containing only carbon
and helium. After the inclusion of oxygen and the line-blanketing by
iron-group elements, \citet{Barniske06} re-analyzed a few of these WC
stars and found that a carbon mass fraction of 40\% appears to be
typical.

  The present study now covers for the first time a large part of the Galactic
sample of WC stars of all subtypes. The analyses comprise 40 WC stars of
subtypes WC4 to WC9. In addition, we consider six stars of the WN/WC
transition type and two WO stars. { The spectral analyses presented
in this work are appropriate for single-star spectra. A few binaries are 
included in order to demonstrate the bias of the obtained parameters if the
composite nature of their spectra is neglected.} 

  The empirical stellar parameters and mass-loss rates of the Galactic WC
stars established by our analyses provide the empirical basis for
studies of (i) the role of WC stars in the evolution of massive stars,
(ii) the wind-driving mechanisms, and (iii) the feedback of WC stars as
input to the modeling of the chemical and dynamical evolution of galaxies.  

  In the next section, we introduce our sample of stars and the
observational data used in this work. The model calculations are
characterized in Sect.\,\ref{sec:grid}. In Sect.\,\ref{sec:method}, we
describe the analyses, and in Sect.\,\ref{sec:results} we present the
results and comment on individual stars. Sect.\,\ref{sec:evolution} then
discusses the results with regard to stellar evolution, before arriving at
our conclusions in the final section (Sect.\,\ref{sec:conclusions}).

\section{Observational data}
  \label{sec:data}

  The current WR catalog from \citet{vdH2001,vdH2006} lists 298 WR stars
in our Galaxy, of which 113 are of WC type.  However, most of the
WR  stars that have been added to the catalog since its previous
release \citep{vdH2001} were discovered in infrared observations,
and suffer from high interstellar extinction in the visual range. Since
our analyses are based on optical spectra, our sample is basically
restricted to the list of \citet{vdH2001}.
  
  A significant fraction of the WR stars are known or suspected to have
a binary companion, mostly of O type. Some of the stars analyzed by
\cite{KH95}, for instance, have since been found to be binaries. In
WR+O binaries, the spectrum is composed of both components. Often the 
O-star's contribution even dominates the overall spectrum. The decomposition 
of these spectra is beyond the scope of this paper. Nevertheless, we include a
few of these composite spectra in our study and analyze them as if they
were single WR stars, to illustrate the effects on the results.
  
  A visible effect of an O-star companion is the so-called ``diluted
emission lines" (d.e.l.). When the light of a companion contributes
significantly to the spectral continuum, the emission lines appear
weaker in the normalized spectrum, hence are called ``diluted". However,
we do not \emph{a priori} judge a WC star as a
binary on the basis of apparently diluted emission lines alone. The
proof of binarity can only be derived from radial-velocity variations or an
eclipse light curve. The d.e.l.\ binary criterion is based on the
assumption that stars of the same subtype have similarly strong emission
lines. For WN stars, \citet{Adriane06} demonstrated that this is not
the case. However, we see below that the WC sequence has rather
uniform spectra, making the d.e.l.\ criterion a valid binary indicator.

  For a few putatively single, visually un-obscured WC stars, we 
have no spectra at our disposal, which leaves a sample of 56 stars for
our study. In addition to the proper WC stars, our sample comprises two
WO stars, five WN/WC transition types, and one WC/WN star (see
Table\,\ref{tab:wcsample}).


\begin{table*}
  \caption{Galactic WC stars, not visually obscured \citep{vdH2001}}
  \label{tab:wcsample}
  \centering

\begin{tabular}{l l c l c l l c l}
\hline 
\hline 
WR  & Subtype\tablefootmark{1} 
                 & Figure\tablefootmark{2}    
                                     & Remarks\tablefootmark{3}     
                                                           & \hfill \rule[0mm]{0mm}{3mm}
                                                             & WR    & Subtype\tablefootmark{1} 
                                                                                   & Figure\tablefootmark{2}    
                                                                                                       & Remarks\tablefootmark{3} \\
\hline 
4   &   WC5      & B.1 &                     & &       &             &                   &            \\
5   &  WC6       & B.2 &                     & & 75a   &   WC9       &  -                & no spectra \\
7a  & WN4h/WC    & -                 & no spectra          & & 76    &   WC9d      &  -                & no spectra \\
8   & WN7/WCE    & B.3 & binary WN+WC?       & & 77    &  WC8+OB     &  -                & binary     \\
9   &  WC5+O7    & -                 & binary              & & 79    & WC7+O5-8    &  -                & binary     \\
11  & WC8+O7.5   & -                 & binary              & & 80    &   WC9d      & B.27 &            \\
13  &  WC7       & B.4 &                     & & 81    &  WC9        & B.28 &            \\
14  &  WC7+?     & B.5 & binary? no d.e.l.   & & 86    & WC7+B0I-III & B.29 & binary: pseudo fit\tablefootmark{5} \\
15  &  WC6       & B.6 &                     & & 88    &  WC9+?      & B.30 & WN+WC or WN/WC9\tablefootmark{6} \\     
17  &  WC5       & B.7 &                     & & 90    &  WC7        & B.31 &            \\
19  & WC4pd+O9.6 & -                 & binary              & & 92    &  WC9        & B.32 &            \\
20a & WN7:h/WC   & -                 & now WN6ha+WN6ha\tablefootmark{4} 
                                                           & & 93    & WC7+O7-9    &  -                & binary     \\
23  &  WC6       & B.8 &                     & & 95    &  WC9d       & B.33 &            \\
26  & WN7/WCE    & B.9 &                     & & 96    &  WC9d       &   -               & no spectra \\
27  &  WC6       & B.10 &                     & & 98    &  WN8/WC7    & B.34 & now WN7/WC+O8-9\tablefootmark{7}: pseudo fit \\
30  & WC6+O6-8   &  -                & binary              & & 98a   & WC8-9vd+?   &   -               & no spectra \\
30a & WO4+O5-5.5 &  -                & binary              & & 101   &  WC8        &   -               & no spectra \\
31c & WC6+OB     &  -                & binary              & & 102   &  WO2        & B.35 &            \\
32  & WC5+OB     &  -                & binary              & & 103   &  WC9d       & B.36 &            \\
33  &  WC5       & B.11 &                     & & 104   & WC9d+B0.5V  & B.37 & binary: pseudo fit\tablefootmark{5} \\
38  &   WC4      & B.12 & binary?             & & 106   &  WC9d       & B.38 &            \\
38b & WC7+OB     & -                 & binary              & & 107a  &  WC6        &   -               & no spectra \\
39  & WC7+OB     & B.13 & binary: pseudo fit\tablefootmark{5}  
                                                           & & 111   &  WC5        & B.39 &            \\
41  & WC5+OB     & -                 & binary              & & 112   & WC9d+OB     &   -               & binary     \\
42  &  WC7+O7V   & -                 & binary              & & 113   & WC8d+O8-9IV & B.40 & binary: pseudo fit\tablefootmark{5} \\ 
45  &  WC6       & B.14 &                     & & 114   &  WC5        & B.41 &            \\ 
47c &   WC5      & -                 & no spectra          & & 117   &  WC9d       & B.42 &            \\ 
48  & WC6+O9.5   & -                 & binary              & & 118   &  WC9d       &   -               & no spectra \\ 
48a &   WC9d     & -                 & no spectra          & & 119   &  WC9d       & B.43 &            \\ 
48b &   WC8ed    & -                 & no spectra          & & 121   &  WC9d       & B.44 &            \\  
50  &  WC7+OB    & B.15 & binary: pseudo fit\tablefootmark{5}  
                                                           & & 125   & WC7ed+O9III & B.45 & binary: pseudo fit\tablefootmark{5} \\ 
52  &   WC4      & B.16 &                     & & 126   &  WC5/WN     & B.46 &            \\
53  &  WC8d      & B.17 &                     & & 132   &  WC6        & B.47 &            \\
56  &   WC7      & B.18 &                     & & 135   &  WC8        & B.48 &            \\
57  &  WC8       & B.19 &                     & & 137   & WC7pd+O9    & B.49 & binary: pseudo fit\tablefootmark{5} \\
58  & WN4/WCE    & B.20 &                     & & 140   & WC7pd+O4-5  &   -               & binary     \\
59  &  WC9d+OB?  & B.21 & binary?             & & 142   &  WO2        & B.50 &            \\
60  &   WC8      & B.22 &                     & & 143   & WC4+Be      & B.51 & binary: pseudo fit\tablefootmark{5} \\
64  &  WC7       & B.23 &                     & & 144   &  WC4        & B.52 &            \\
65  & WC9d+OB?   & B.24 & binary?             & & 145   & WN7/WCE     & B.53 &            \\
68  &  WC7       & B.25 &                     & & 146   & WC6+O8      & B.54 & binary: pseudo fit\tablefootmark{5} \\
69  & WC9d+OB?   & B.26 & binary?             & & 150   &  WC5        & B.55 &            \\
70  & WC9vd+B0I  & -                 & binary              & & 153ab & WN6/WCE+O6I &   -               & binary     \\
73  &  WC9d      &  -                & no spectra          & & 154   &  WC6        & B.56 &            \\
\hline 
\end{tabular}
\tablefoot{
  \tablefoottext{1}{Classification from \citet{vdH2001}}
  \tablefoottext{2}{Spectral fits, shown in the online appendix}
  \tablefoottext{3}{Further comments to the individual stars are given in Sect.\,\ref{appsec:starcomments}}
  \tablefoottext{4}{\citet{WR20aPaper}}
  \tablefoottext{5}{Known binary fitted as if it were a single star (as discussed in Sect.\,\ref{sec:data})}
  \tablefoottext{6}{\citet{WC9cwb}}
  \tablefoottext{7}{\citet{WR98bin}}
}   
\end{table*}

  
  For the analyses, we collected optical spectra from both \citet{SpecTM87}
and our own observations performed at the Calar Alto observatory or ESO. 
The optical spectra typically cover the wavelength range 
3300-7300\,\AA. In some cases however, if a star e.g. is not bright enough in a
certain wavelength regime or there were technical difficulties during the
observation, only a part of this range is covered. The typical spectral 
resolution is around 10\,\AA\ \citep{SpecTM87}.
For the ultraviolet range, IUE spectra were retrieved from the INES archive.
They cover the range from 1150\,\AA\ to 3200\,\AA\ with a typical resolution of 
around 6\,\AA. Infrared photometry (J, H, K) was taken from the 2MASS catalog
\citep{2MASS}, and in two cases (WR\,52, WR\,142) was augmented by Spitzer IRS
observations. For some stars, we used additional optical narrow-band photometry
by \citet{LS84}, MSX infrared photometry \citep{MSX}, and/or Spitzer spectra 
from \citet{SASS}.

\section{Model grid}
  \label{sec:grid}

  To reproduce the observed spectra, we used the Potsdam
Wolf-Rayet (PoWR) code for expanding stellar atmospheres. The model
basics are described in \citet{GKH02} and \citet{HG04}. The non-LTE
radiative transfer is calculated for a spherically symmetric and
stationary outflow with a pre-described velocity field. In the
supersonic part, we adopt the usual $\beta$-law with the terminal
velocity $\varv_{\infty}$. The exponent $\beta$ is set to 1.0 throughout
this study, except for the two WO stars for which $\beta = 0.8$ gave
slightly better fits. In the subsonic part, the velocity field is defined
by the requirement that the density stratification approaches the
hydrostatic limit, as described by \citet{HS87}. Since the mass-loss
rate $\dot{M}$\ and the velocity field are treated as free parameters,
the models are not hydrodynamically consistent.  
  
  Wind inhomogeneities are taken into account in the ``microclumping''
approximation, assuming that individual ``clumps'' are optically thin at
all frequencies, and that the interclump medium is void. The clumping
factor $D$ denotes the density within the clumps, relative to the
homogeneous model with the same $\dot{M}$, and is assumed to be constant
throughout the atmosphere in this study.

Complex model atoms are taken into account, as summarized in
Table\,\ref{tab:grids}. Iron-group elements are treated in the
superlevel approximation \citep{GKH02}.

  The inner boundary of our model atmospheres is located at a radial
optical depth of $\tau_{\mathrm{Ross}} = 20$, which we define to correspond
to the stellar radius $R_{*}$. We define the ``stellar temperature'' $T_{*}$
as the effective temperature related to that radius, via Stefan-Boltzmann's
law
  \begin{equation}
    \label{eq:lrt}
    L = 4 \pi R_{*}^2 \sigma_{\mathrm{SB}} T_{*}^4
  \end{equation}
and the stellar luminosity $L$.  

  As \citet{SHW89} found out, stellar winds produce very similar
emission-line spectra when they agree in a certain combination of the
mass-loss rate and other parameters. To exploit this scaling invariance,
they introduced the so-called ``transformed radius''
\begin{equation}
  \label{eq:rt}
 R_{\mathrm{t}} = R_{*} \left[ \frac{\varv_{\infty}}{2500\,\mathrm{km/s}} 
 \left/ \frac{\dot{M} \sqrt{D} }{10^{-4} M_{\odot}/\mathrm{yr}} \right. 
 \right]^{\frac{2}{3}}
\end{equation} 
\citep[the clumping factor $D$ has been incorporated in this definition
by][]{HK98}.
   
  The name ``transformed radius", historically coined by \citet{SHW89},
is actually misleading since $R_{\mathrm{t}}$ does not have the meaning
(although the units) of a radius. A more appropriate quantity to consider
would be $R_{\mathrm{t}}^{-3}$, which might be called a ``normalized emission
measure". Since this quantity is proportional to the volume integral of the 
density squared, divided by the stellar surface, $R_{\mathrm{t}}^{-3}$ scales
with the emission from recombination lines normalized to the continuum.
This explains why different combinations of $R_{*}$, $\varv_{\infty}$,
and mass-loss rate $\dot{M}$ result in approximately the same normalized
WR emission-line strengths as long as $R_{\mathrm{t}}$ (or
$R_{\mathrm{t}}^{-3}$) is kept at the same value.

  Wolf-Rayet model spectra are most sensitive to $T_{*}$ and
$R_{\mathrm{t}}$. When finding systematically the model that provides the 
closest fit to a given observation, it is therefore most convenient to rely 
on \emph{model grids} in which these two parameters are varied. The other
parameters, namely the chemical composition and the terminal wind
velocity, are kept constant within one model grid.
  
  We established one large model grid for most of the WC
parameter space, and two smaller grids for the WC9 domain. The grid
resolution is 0.05 in $\log\,T_{*}$ and 0.1 in $\log\,R_{\mathrm{t}}$.
For the chemical composition, we assumed mass fractions of 55\% helium,
40\% carbon, 5\% oxygen, and 0.16\% iron-group elements (cf.\
Table\,\ref{tab:grids}). To verify that these abundances are adequate to
the observed stars, we also calculated smaller grids with 60\% and 20\%
carbon fractions, respectively.

For the main WC grid, we set the terminal wind velocity to
2000\,km\,s$^{-1}$, which is in the middle of the range actually
observed. The two WC9 grids are calculated for 1000\,km\,s$^{-1}$ and
1600\,km\,s$^{-1}$. Owing to their their lower ionization, the WC9 models
include an extended \ion{C}{ii} model atom.

Throughout this study, the microturbulence velocity is set to 100\,km/s, 
based on the experience that this yields profile shapes similar to the 
observed ones. For the clumping density factor, we adopt $D$ = 10, which 
generally reproduces the observed strength of the electron-scattering wings. 
However, the true degree of the wind inhomogeneity remains the subject of debate
\citep[cf.][]{ClumpingWS2008}. According to Eq.\,(\ref{eq:rt}), the
empirically obtained mass-loss rates scale with the adopted clumping
factor as $\dot{M} \propto D^{-1/2}$.

The luminosity was kept fixed over all our grids at $\log
L/L_{\odot} = 5.3$. According to Eqs.\,(\ref{eq:lrt}) and (\ref{eq:rt}),
the scaling of the model to a different luminosity implies a scaling of
the mass-loss rate as $\dot{M} \propto L^{3/4}$. In a strict sense, the
scaling invariance between models of the same $R_\mathrm{t}$ is only
approximate, but accurate enough to cover the range of luminosities in
our sample.
  
\begin{figure}[!bth]
  \resizebox{\hsize}{!}{\includegraphics{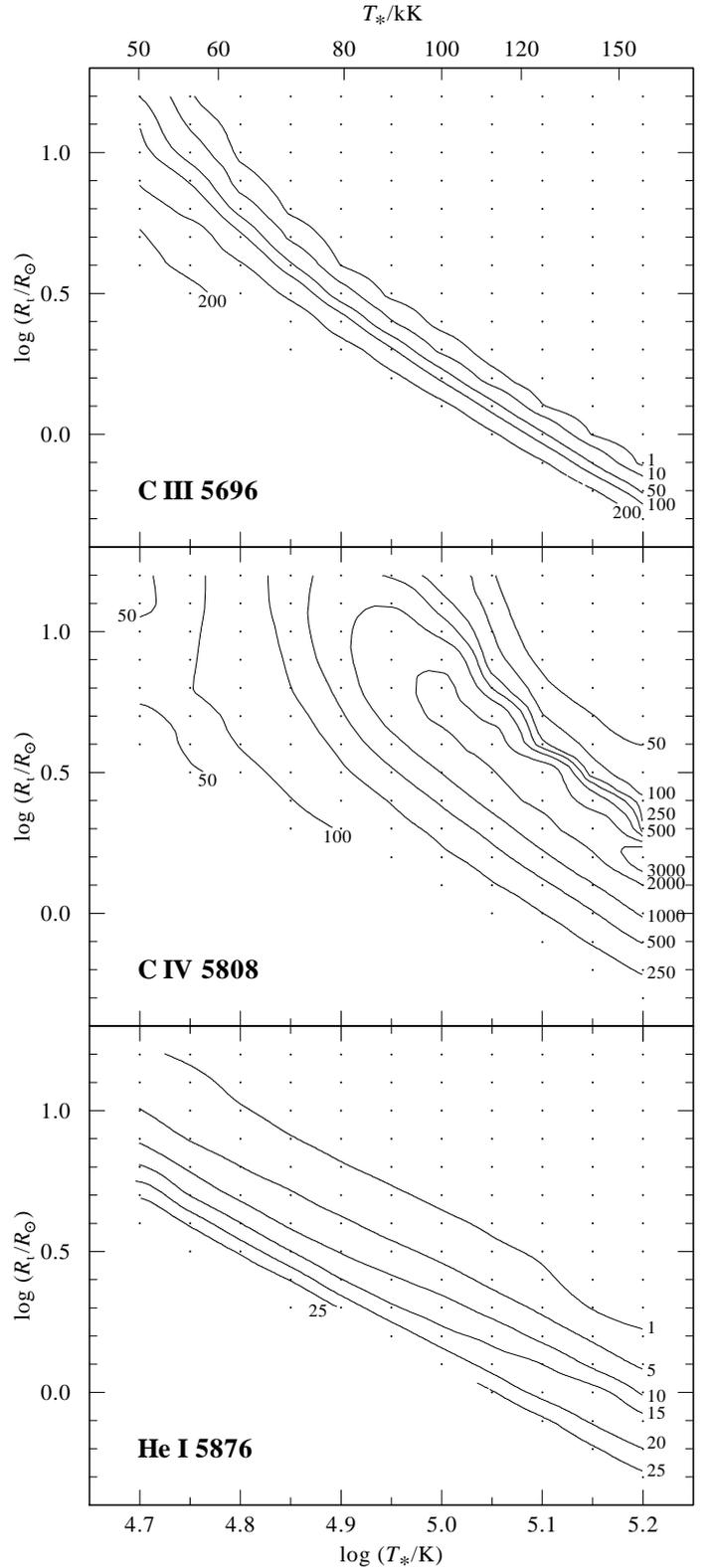}}
  \caption{Contours of constant equivalent widths of the WC emission lines \ion{C}{iii}\,5696\,\AA, \ion{C}{iv}\,5808\,\AA\ and \ion{He}{i}\,5876\,\AA. The small dots represent models of the WC grid. Labels give absolute value of the equivalent width in \AA.}
  \label{fig:isocarbonhelium}
\end{figure} 

\begin{figure}[bth]
  \resizebox{\hsize}{!}{\includegraphics{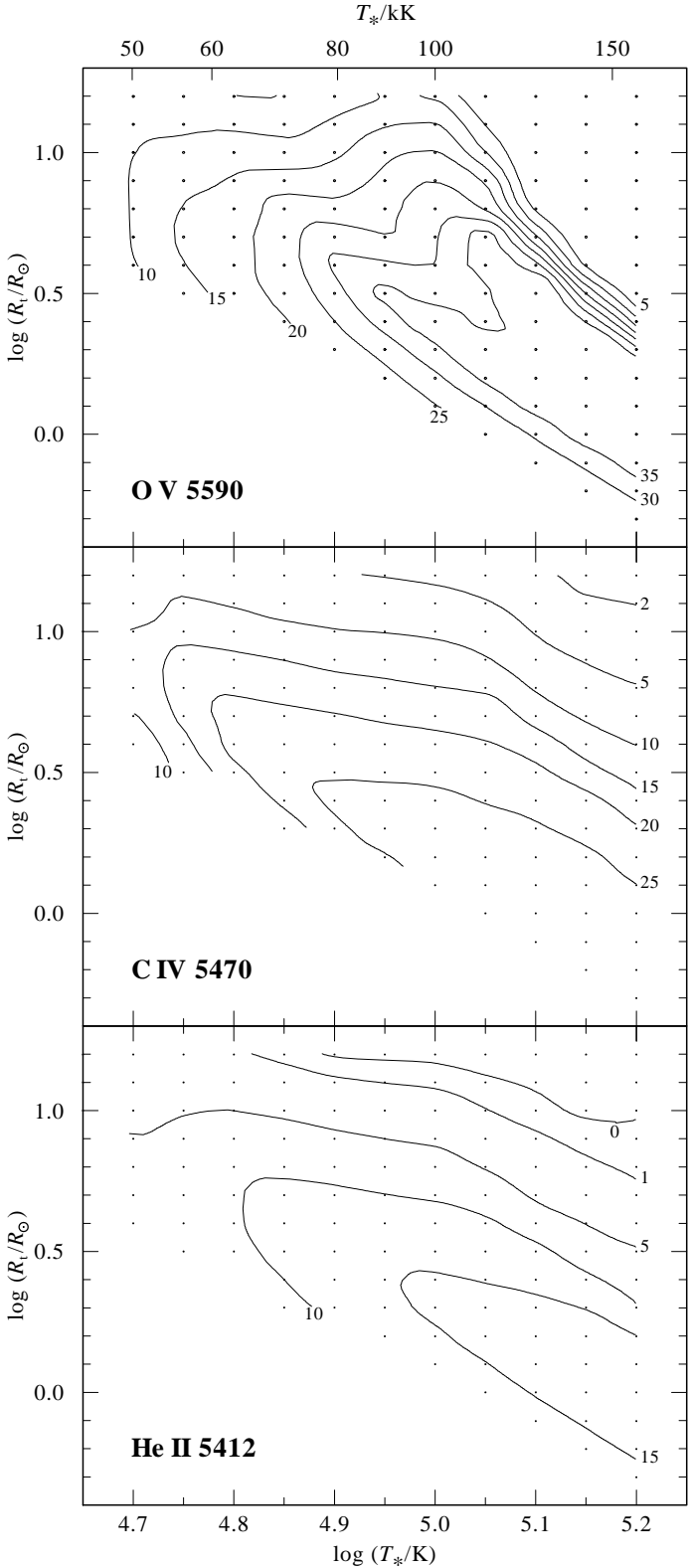}}
  \caption{Same as Fig.\,\ref{fig:isocarbonhelium}, but for the lines \ion{O}{v}\,5590\,\AA, \ion{C}{iv}\,5470\,\AA\ and \ion{He}{ii}\,5412\,\AA.}
  \label{fig:isooxygendiagpair}
\end{figure} 

  Figures \ref{fig:isocarbonhelium} and \ref{fig:isooxygendiagpair}
display contour plots for the strength of the brightest WC emission lines.
The small dots represent the calculated models of the standard WC grid.
Each contour refers to a certain equivalent width, labeled with
$-W_{\lambda}$ in \AA. The ratio of the lines
\ion{C}{iii}\,5696\,\AA\ to \ion{C}{iv}\,5808\,\AA\ is used as the
primary criterion for the WC subtype classification
\citep{WCClassCrowther}. For early subtypes, \ion{O}{v}\,5590\,\AA\ is
also considered \citep{vdH2001}. 
  
  For the first time, we apply the PoWR code when analyzing spectra of the
late subtype WC9. Most of these stars show an excess of infrared
emission that is attributed to circumstellar dust, as denoted by the
letter ``d'' appended to their classification, WC9d. This circumstellar
dust usually leads to strong reddening of the stellar radiation,
inhibiting  their observation in the UV. The optical spectrum, however,
does not suffer much from the dust emission and absorption, and can be
analyzed with the help of our models together with the non-dusty WC9
stars.

  The WC9 stars have a lower stellar temperature than the earlier WC
subtypes, which is evident from the lower ions that are visible in their
spectra. We therefore calculated special, small grids of models in the
WC9 parameter domain, including here the extended \ion{C}{ii} model atom. 
Compared to the earlier subtypes, the WC9 stars have slower winds, as 
indicated by the smaller width of their lines. We therefore calculated 
one WC9 grid with $\varv_{\infty}\,=\,1600\,$km/s and another with only
$\varv_{\infty}\,=\,1000\,$km/s.

  The two WO-type stars in our sample turned out to be very hot ($T_{*}
\approx 200\,$kK). We therefore had to include higher ions of the
iron-group elements, up to \ion{Fe}{xv}, in their models. 
The extremely broad emission lines indicate wind velocities as high as
$\varv_{\infty} \approx 5000$\,km/s, which is also much higher than for
the WC stars. Hence, a special set of model calculations hat to be used
to analyze the WO-type stars.
 
\begin{table} 
  \caption{Model grid parameters}
  \label{tab:grids}
  \centering
  \begin{tabular}{l c c c}
  \hline\hline
    				  \rule[0mm]{0mm}{3mm}    		&	WC grid & WC9 grids \\
  \hline
     $X_{\mathrm{He}}$ \rule[0mm]{0mm}{3mm} &	\multicolumn{2}{c}{55\%} \\
     $X_{\mathrm{C}}$		                    &	\multicolumn{2}{c}{ 40\% }  \\
     $X_{\mathrm{O}}$		                    &	\multicolumn{2}{c}{  5\% }  \\
     $X_{\mathrm{Fe}}$\tablefootmark{a}		  & \multicolumn{2}{c}{0.16\%}  \\
     $\log L/L_{\odot}$									  	& \multicolumn{2}{c}{ 5.3  }  \\
     $\varv_{\infty}$\,/\,km\,s$^{-1}$      &	 2000   & 1000/1600\tablefootmark{b} \\
     $D$                           & \multicolumn{2}{c}{   10 }  \\[2mm]
     
     \multicolumn{2}{l}{\textit{Number of Levels}} \\
     \ion{He}{i}                            &	\multicolumn{2}{c}{		17 }	\\
     \ion{He}{ii}                           &	\multicolumn{2}{c}{		16 }	\\
     \ion{He}{iii}                          &	\multicolumn{2}{c}{	  1	 }	\\
     
     \ion{C}{ii}										  			&		  	3      &       29	    \\
     \ion{C}{iii}										  			&	\multicolumn{2}{c}{		40 }	\\
     \ion{C}{iv}										  			&	\multicolumn{2}{c}{		19 }	\\
     \ion{C}{v} 										  			&	\multicolumn{2}{c}{		1  }	\\

     \ion{O}{ii} 									  				&	\multicolumn{2}{c}{		3  } 	\\
     \ion{O}{iii} 							  					&	\multicolumn{2}{c}{		33 }	\\
     \ion{O}{iv} 										  			&	\multicolumn{2}{c}{		25  }	\\
     \ion{O}{v} 											  		&	\multicolumn{2}{c}{		36  }	\\
     \ion{O}{vi} 												  	&	\multicolumn{2}{c}{		15  }	\\
     \ion{O}{vii}                           &	\multicolumn{2}{c}{		1   }	\\

     \ion{Fe}{iii}-\textsc{x}\tablefootmark{a}	
                                            & \multicolumn{2}{c}{72\,superlevels}	\\
  \hline
  \end{tabular}
  \tablefoot{
  	\tablefoottext{a}{Generic element, representing also Sc, Ti, V, Cr, Mn, Co, and Ni, 
  	                  with relative abundances to Fe as described in \citet{GKH02}}
  	\tablefoottext{b}{We calculated two WC9 grids with different velocities}
  }
\end{table}
 
  The synthetic spectra of the main WC grid and the WC9 sub-grids will
be made available from our
website\footnote{\texttt{http://www.astro.physik.uni-potsdam.de/PoWR.html}}
simultaneously with the release of this paper.
  
\section{The analysis method}
  \label{sec:method}
  
  The stars are analyzed by finding the PoWR model that most closely fits
the observations. The first part of the fit procedure refers to the
normalized line spectrum, and the second part to the fit of the spectral
energy distribution (SED). Since WC spectra are full of lines, a rectification
``by eye'' can be ambiguous. Fortunately, most of the spectra we used are
calibrated in absolute fluxes. We therefore rectify the spectra by
dividing them by the \emph{model continuum}. Consequently, the two steps
(a) line fit and (b) SED fit are actually coupled and have to be
iterated.  
  
  The WC stars are divided into subtypes based on the ratios of the
equivalent widths of certain emission lines. The standard classification
system was summarized by \citet{vdH2001} and uses the ratio of
\ion{C}{iv} to \ion{C}{iii} as a primary criterion. The main
classification lines are \ion{C}{iv}\,5808\,\AA\ and
\ion{C}{iii}\,5696\,\AA, where \ion{C}{iv}\,5808\,\AA\ is either one of two
lines, together with the \ion{C}{iii}/\ion{C}{iv}/\ion{He}{ii}-blend around
4650\,\AA, that are extremely strong in the optical spectrum of early WC
subtypes.
  
  One of our criteria for choosing the best-fit model is to ensure in
particular that we can fit these strongest lines. However, the enormous peak 
heights of in particular the earlier subtypes can only be reached
in the models at the price that other emission lines become stronger
than observed. Thus, we often had to make some compromise with
respect to the overall fit quality, and also paid attention to other
significant lines such as \ion{O}{v}\,5590\,\AA, the ``diagnostic line
pair'' (\ion{He}{ii}\,5412\,\AA\ and \ion{C}{iv}\,5470\,\AA), and the
carbon lines in the UV. For some stars in our sample, the red part of the
optical spectrum was not available, and our fit was restricted to the
4650\,\AA-blend and neighboring lines such as \ion{C}{iv}\,4441\,\AA\ and
sometimes \ion{O}{iv}\,3412\,\AA.

\begin{figure*}
  \centering
  \includegraphics[angle=-90,width=17cm]{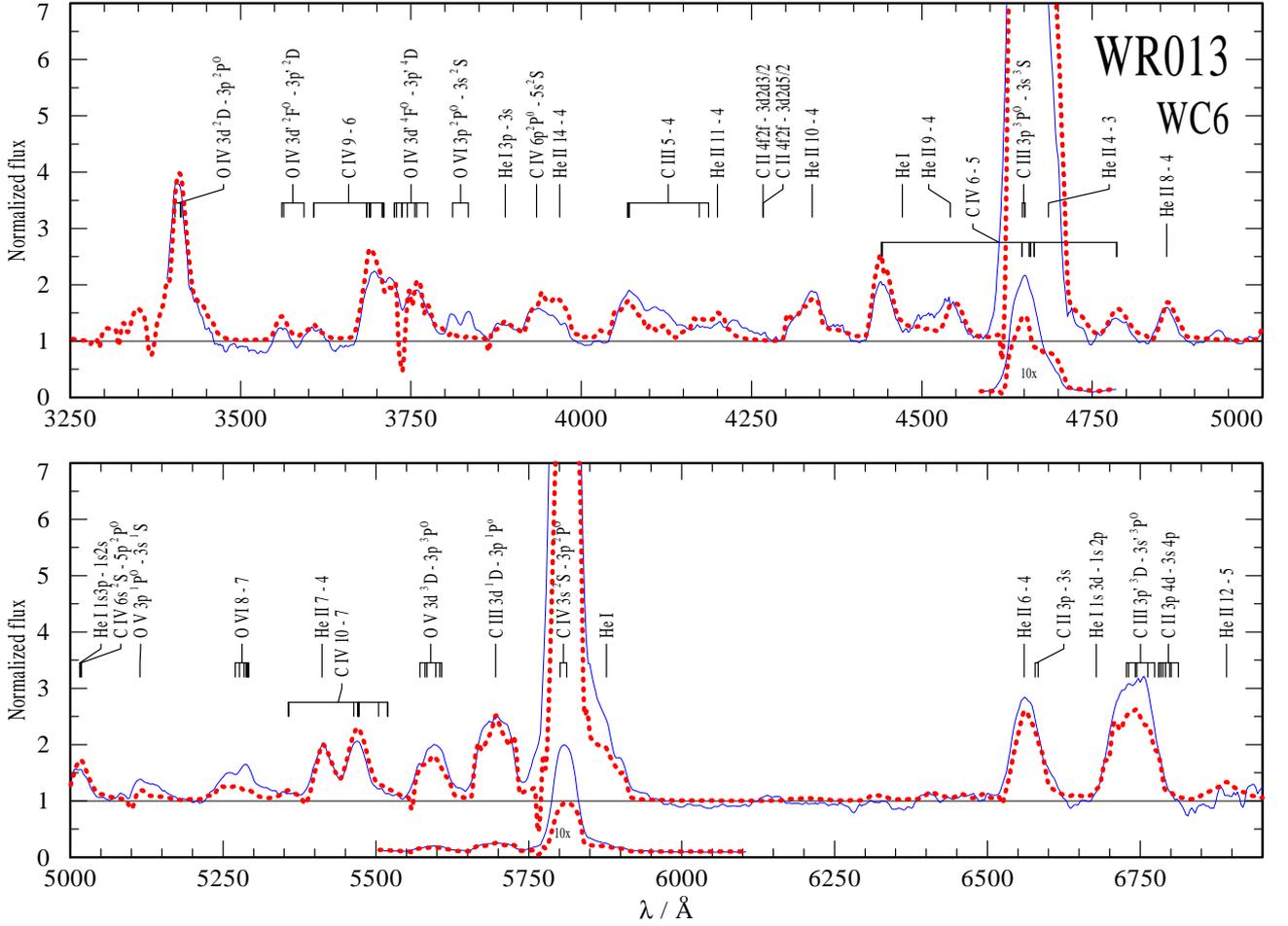}
  \caption{Optical spectrum of the WC6 star WR\,13. The solid thin line 
    is the observed spectrum, the thick dotted line represents the 
    best-fit WC grid model. The primary model parameters are 
    $T_{*}\,=\,79.4\,$kK and $\log R_{\mathrm{t}}/R_{\odot}\,=\,0.5$. 
    The same model fits to most of the Galactic WC6 single stars, although 
    it tends to underestimate the peak heights of \ion{C}{iv}\,5808\,\AA\ 
    and the \ion{C}{iii}/\ion{C}{iv} blend around 4650\,\AA.}
  \label{fig:wr13opt}
\end{figure*}

  Hence, the line fit yields the stellar temperature $T_{*}$ and the 
``transformed radius'' $R_{\mathrm{t}}$ as main parameters, together with 
constraints on the terminal wind velocity $\varv_{\infty}$ and the 
chemical composition. Fig.\,\ref{fig:wr13opt} shows as an
example the observed optical spectrum of the WC6 star WR\,13, together
with the synthetic spectrum from the best-fit model.   
  
\begin{figure*}
  \centering
  \includegraphics[angle=-90,width=17cm]{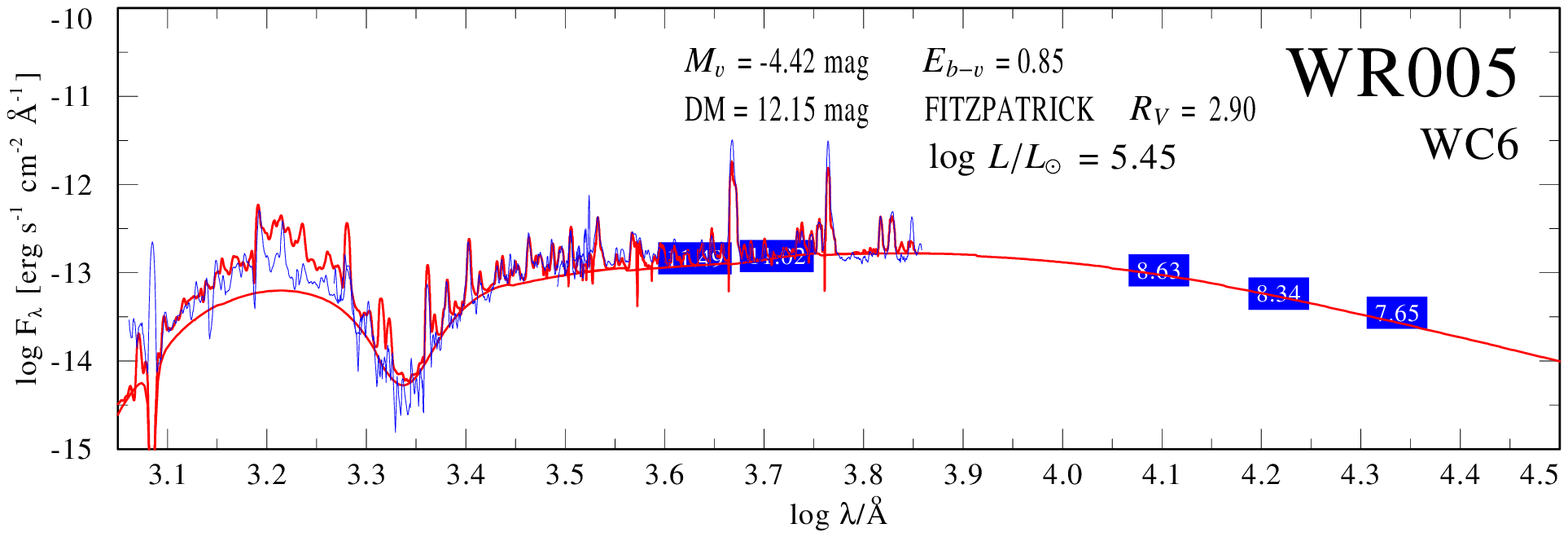}
  \caption{Spectral energy distribution (SED) for the WC6 star WR\,5. 
    Flux-calibrated observations (blue thin and noisy line) are from IUE in
    the UV, \citet{SpecTM87} in the optical, and 2MASS ($J$, $H$, and $K$
    photometry) in the near-infrared. The synthetic spectrum is plotted as a
    red thick line. The continuum-only model flux is also shown for
    comparison. The model that fits the normalized line spectrum best has
    been selected from the WC standard grid ($T_{*}\,=\,79.4\,$kK, $\log
    R_{\mathrm{t}}/R_{\odot}\,=\,0.5$).}
  \label{fig:wr5sed}
\end{figure*}  
  
  The PoWR models do not only provide a normalized spectrum but also the 
absolute flux values of a star that can be used to fit the SED if 
flux-calibrated spectra and/or photometric 
values are available. The dilution of the model flux depends on the 
distance and the reddening. For stars with a known distance modulus (D.M.), 
we can obtain the absolute magnitude $M_{\varv}$ by  
  \begin{equation}
    \label{eq:Mv}
  	M_{\varv} = m_{\varv} - \mathrm{D.M.} - A_{\varv} ~.
  \end{equation}  
We note that the color indices ($\varv$ and $b$) refer to the monochromatic
magnitudes defined by \citet{Smith1968}. The value of $A_{\varv}$
is provided by the SED fit where the reddening parameter $E_{b-\varv}$
is a free parameter, as well as the detailed reddening law. Whenever
adequate, we applied the reddening law from \citet{Seaton}, which implies a
fixed ratio of extinction $A_{V}$ to color excess $E_{B-V}$ of 
$R_{V} = 3.1$. If this turned out to be insufficient in reproducing the SED, 
we used the laws of \citet{Cardelli} or \citet{Fitzpatrick}, which
treat $R_V$ as a free parameter, and adjusted the latter to optimize the
SED fit. (The reddening law and $R_V$ are both indicated in the online material 
plots.) If mid-IR spectra were available for the SED fit, we always chose the 
reddening law of \citet{Fitzpatrick}. 

An example SED fit is shown in Fig.\,\ref{fig:wr5sed}. As the distance for 
this star is not known, an absolute magnitude $M_{\varv} = -4.42\,$mag was adopted
from our subtype calibration. With the help of Fig.\,\ref{fig:wr5sed}, we found that 
the luminosity of the grid model must be scaled to $\log L/L_{\odot} = 5.45$ in order to 
fit the observed SED. Simultaneously, the color excess $E_{b-\varv}$ is also properly 
adjusted, together with the choice of the reddening law (``\textsc{cardelli}'') and its 
parameter $R_{V}$. The fit implies a distance modulus of D.M.\ = 12.15\,mag.

  {The color excess $E_{b-\varv}$ in the Smith system is related to
the more common $E_{B-V}$ value from the Johnson system by $E_{B-V} = 1.21 E_{b-\varv}$,
while the extinction $A_{\varv}$ is given by
  \begin{equation}
    \label{eq:Av}
    A_{\varv} = \left( 1.21 R_{V} + 0.36 \right) E_{b-\varv}
  \end{equation}  
  as described by \citet{LS84}.}  

\begin{table}
  \caption{Calibration of absolute magnitudes for the different WC 
    subtypes, as derived from stars with distances known from their 
    cluster or association membership (cf.\ Fig.\,\ref{fig:wcmvcalib})}
  \label{tab:wcmvcalib}

  \centering
  \begin{tabular}{lc}
      \hline
      \hline
Subtype  \rule[0mm]{0mm}{3mm} & \multicolumn{1}{c}{$M_{\varv}$ [mag]} \\
      \hline  
      WC4  \rule[0mm]{0mm}{3mm} &  $-3.34$  \\
      WC5  &  $-4.12$  \\
      WC6  &  $-4.42$  \\
      WC7  &  $-4.18$  \\
      WC8  &  $-4.48$  \\
      WC9  &  $-5.13$  \\[2mm]

      WN4/WCE &      $-3.84$\tablefootmark{a}     \\
      WN7/WCE &      $-5.67$\tablefootmark{a}     \\
    \hline
  \end{tabular}
  \tablefoot{
  	\tablefoottext{a}{Adopted WN calibration from \citet{Adriane06}}
  }
\end{table}


  In general, it is hard to determine the distances of individual stars
in our Galaxy, hence only for a subset of our sample are the distances 
known based on their membership to open clusters or associations 
(see Sect.\,\ref{appsec:starcomments}). For
all other stars we follow the classical approach of assuming that all
stars of the same subtype have the same absolute visual brightness
$M_{\varv}$.

\begin{figure}[ht]
  \resizebox{\hsize}{!}{\includegraphics[angle=-90]{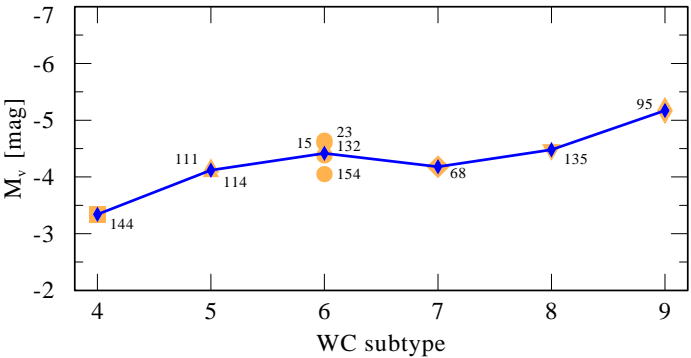}}
  \caption{$M_{\varv}$ calibration for the different WC subtypes, using
    stars for which the distance is known from their membership to 
    an open cluster or association (labels: WR number). 
    The symbols connected by the dark line 
    show the calculated average values for $M_{\varv}$. The obtained 
    values are listed in Table\,\ref{tab:wcmvcalib}.}
  \label{fig:wcmvcalib}
\end{figure}

  From the WC stars of our sample with known distances, the
$M_{\varv}$ values are shown in Fig.\,\ref{fig:wcmvcalib}. The average
relation is also indicated and tabulated in Table\,\ref{tab:wcmvcalib}.
It is unclear whether the assumption of a unique absolute brightness
within each subtype is justified. The scatter in the $M_{\varv}$ values
within one subtype in Fig.\,\ref{fig:wcmvcalib} may only reflect the
errors in the adopted stellar distances, but could also by partly
intrinsic to the stars. Within these uncertainties, the resulting 
$M_{\varv}$ values increase towards later subtypes. This is similar
to the corresponding subtype calibration of the hydrogen-free 
WN stars \citep{Adriane06}. We note that even the same visual brightness 
would still imply a sharp dependence of the luminosity on subtype, because 
of their different bolometric corrections (see below).

  For the two WN/WC stars without known distances, we adopt the $M_\varv$
calibration for the corresponding hydrogen-free WN subtype of
\citet{Adriane06}, because their chemical composition is more similar to 
WN than WC stars.


\section{Results}
  \label{sec:results}

\subsection{Stellar parameters}
  \label{subsec:param}
  
The results of the analyses are compiled in Table\,\ref{tab:wcpar} for our
whole sample, now sorted by spectral subtype. The spectral fits and comments
on individual stars are given in the appendix (online version only, 
see Sect.\,\ref{appsec:specfits} and \ref{appsec:starcomments}).

\begin{figure*}[bth]
  \centering
  \includegraphics[angle=-90,width=17cm]{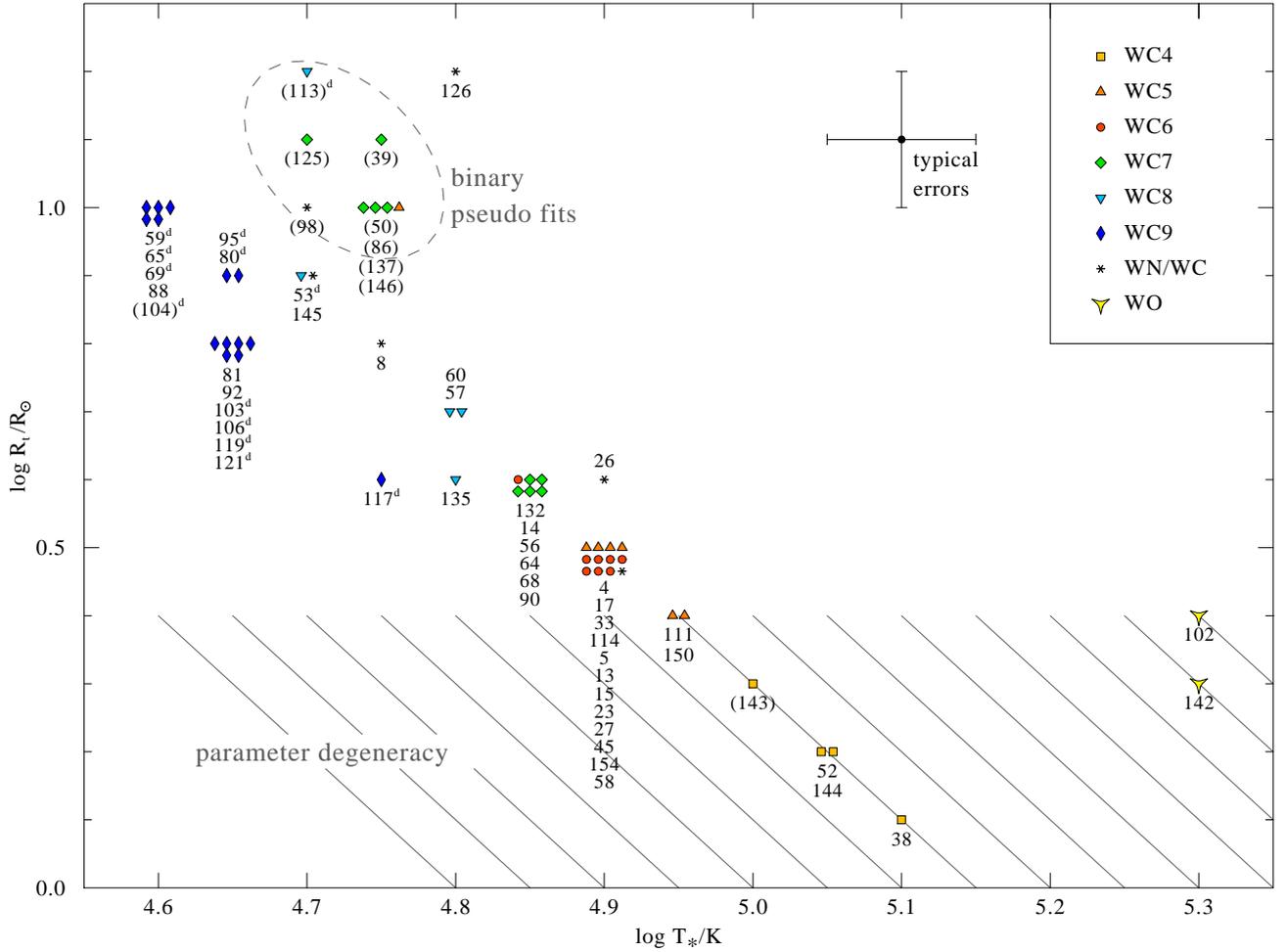}
  \caption{The results of the WC analyses in the $\log\,T_{*}$-$\log
    R_{\mathrm{t}}$-plane, which corresponds to the basic model
    parameters. The WC stars form a one-dimensional sequence with a slight
    offset for the WC9 stars. Stars appearing above the sequence turn out to
    be binaries. The symbol shapes indicate the different WC subtypes. The
    numbers next to the symbols are the WR numbers according to
    \citet{vdH2001}.
    Binary systems analyzed as a pseudo fit have WR numbers in brackets.
    Stars with persistent dust emission are marked with a superscript ``d".
    {The thin grey lines in the lower part indicate the thick wind regime 
    where stars can be shifted along these lines without significant changes
    in the synthetic spectrum.}}
  \label{fig:wcstats}
\end{figure*}

  We begin by discussing the results from the line fit, chiefly the
stellar temperature and the transformed radius. Figure\,\ref{fig:wcstats} shows the
locations of the analyzed stars in the $\log R_{\mathrm{t}}$-$\log
T_{*}$-plane. The striking result visible in that diagram is the clear,
one-dimensional  sequence of the WC subtypes. They are aligned along the
linear relation
\begin{equation}
  \label{eq:wctrend}
  \log R_{\mathrm{t}} \propto -2 \log T_{*}
\end{equation}
with little scatter. The stellar temperature of the WC stars
correlates tightly with their subtype. The two WO stars in our sample, 
however, do not appear to follow an extrapolation of the WC sequence, but 
are much hotter.

  Another group of points in Fig.\,\ref{fig:wcstats} refers to stars that
are known or highly suspected to be binaries with a luminous companion. We 
analyzed their spectra as if they were single stars. The
parameters obtained by these ``pseudo fits'' clearly distinguish the
binaries from the sequence of single WC stars. This independently
confirms the composite nature of their spectra. Thus, the ``diluted
emission line" (d.e.l.) criterion, which is insufficient for WN stars as
emphasized by \citet{Adriane06}, seems to work fine for WC stars when 
distinguishing single stars from binaries. We conclude that those stars that 
follow the WC sequence in Fig.\,\ref{fig:wcstats} do not suffer significantly
from the line dilution effect, i.e.\ they do not have a
luminous companion.

  The tight correlation of the parameters with WC subtype visible in
Fig.\,\ref{fig:wcstats} is now much clearer than in the first 
comprehensive analyses of Galactic WC stars by our group \citep{KH95}. 
Moreover, the subclass of weak-lined WC stars (WC-w) that we had 
introduced in the old paper is no longer necessary: \emph{all} WC-w 
stars have since been identified as binaries.

  Some deviation from the straight alignment of subtypes is seen in
Fig.\,\ref{fig:wcstats} for the WC9 stars. 
The WC9 stars are characterized by the appearance of 
strong \ion{C}{ii}-lines, and are significantly cooler than the 
other WC subtypes. 

  In most of the WC9 spectra, the lines are narrower than in the
simulations of our standard WC grid, which is calculated with a terminal
wind speed of $\varv_{\infty}$\,=\,2000\,km\,s$^{-1}$. We therefore
calculated special grids in this parameter range with
$\varv_{\infty}$\,=\,1600\,km\,s$^{-1}$ and
$\varv_{\infty}$\,=\,1000\,km\,s$^{-1}$.

  The fit quality for the WC9 spectra differs from those of the other subtypes.
While the ``IUE long" UV spectra (1900--3300\,\AA) are usually accurately
reproduced and the ``IUE short" range (1200--2000\,\AA) is hardly met,
the opposite is often true for the few WC9 stars with available UV spectra
(WR\,69, WR\,92, WR\,103).

  In the optical part of the spectrum, most lines are reproduced consistently 
by our models, although the peak heights of \ion{C}{iii}\,4648\,\AA\ and some 
lines between 6500\,\AA\ and 7200\,\AA\ are not recreated. The temperature steps 
in our grid are a bit too coarse to find the optimum fit in some cases, e.g.\ when 
the lines \ion{C}{iv}\,5808\,\AA\ and \ion{He}{i}\,5876\,\AA\ are observed
with almost the same peak heights.

  The WC9 stars of our sample cluster into two distinct groups in the $\log
R_{\mathrm{t}}$-$\log T_{*}$-plane, one cooler group around $T_{*}
\approx$\,40\,kK and $\log\,R_{\mathrm{t}}$\,=\,1.0, which we 
refer to as group I, and another one with higher $T_{*}
\approx$\,45\,kK and a denser wind ($\log\,R_{\mathrm{t}}$\,=\,0.8), which we
call group II. Could it be that group I consists of yet undetected 
binaries, while only the stars in group II are really single stars?

In group I, only  WR\,104 (the ``pinwheel'') is an established binary,
but all others are suspected to be binaries (see appendix
Sect.\,\ref{appsec:starcomments} for corresponding details about WR\,59,
WR\,65, WR\,69, and WR\,88). On the other hand, the WC9d star WR\,119 
from the hotter group II was checked for binarity by \citet{WC9cwb} 
with a negative result. The same holds for WR 117, which is even hotter than 
both groups and closer to the WC8 stars in Fig.\,\ref{fig:wcstats}. Two 
of the three WC9 stars in our sample that do not show dust 
(WR 81, and WR\,92) belong to the hotter group II, which seem to 
consist of the definitive non-binaries. The third one (WR\,88) seems to
be a special case as discussed in Sect.\,\ref{appsec:starcomments}.

  For very dense winds ($\log R_{\mathrm{t}}/R_{\odot} \la 0.4$), the
spectra no longer depend on the two independent parameters $T_{*}$\
and $R_{\mathrm{t}}$, but only on the product $R_{\mathrm{t}} T_{*}^2$
\citep{HGK03}. The physical reason for this approximate parameter
degeneracy is that in very dense winds, \emph{all} radiation (including
the continuum) emerges from layers that expand with nearly the terminal
wind speed. Hence, under these conditions, the mass-loss rate is the only
relevant parameter when comparing models with same luminosity and
$\varv_{\infty}$. Lines of constant $R_{\mathrm{t}} T_{*}^2$ (or,
equivalently, constant $L/\dot{M}^{4/3}$) are indicated by thin parallel
lines in the lower part of Fig.\,\ref{fig:wcstats}. The model spectra in
this range are nearly the same along these lines, hence the
specific combination of $R_{\mathrm{t}}$ and $T_{*}^2$ is not well-constrained. 
In other words, we see only rapidly moving layers of the
atmosphere, while the slower part of the wind is opaque at all
wavelengths, and therefore the radius of the hydrostatic stellar core is not
observable. This degeneracy primarily affects the WC4 stars.

  When available and not in conflict with our spectral fits, terminal
velocities compiled in Table\,\ref{tab:wcpar} were
collected from \citet{Prinja90} and \citet{NiSk2002}. These values had been 
derived from UV spectra, using the blue end of P Cygni line profiles.
Therefore, they might differ from the terminal velocities obtained from 
optical emission lines. In these cases and for all remaining stars
-- mainly those without existing UV observations -- $\varv_{\infty}$ was
roughly estimated by ourselves from the widths of the optical lines. 
We note that the models used for the line fits shown in the Appendix are not 
calculated with the individual wind velocity of the star, but are taken 
from the models grids with fixed $\varv_{\infty}$. The individual values
are inserted, however, when Eq.\,(\ref{eq:rt}) is applied.

  We have so far presented those results derived from fits of the 
normalized line spectra. We now consider parameters which involve 
absolute dimensions, namely $R_{*}$, $\dot{M}$, and $L$, and those required 
to fit the SED in terms of absolute fluxes.

  The color excess $E_{b-\varv}$, for which we use the monochromatic
colors defined by \citet{Smith1968}, of the best SED fit is included in
Table\,\ref{tab:wcpar}. Further details, i.e.\ the applied reddening law
and its $R_V$ parameter if applicable, are printed on the fit plot shown
for each star in the online Appendix.

  As explained above in Sect.\,\ref{sec:method}, the SED fit relies on either
the adopted distance or absolute brightness, as indicated by the
direction of the little arrows between the respective columns in 
Table\,\ref{tab:wcpar}. 

  The basic free parameter of the SED fit was the stellar luminosity. The
models, which were all calculated for a fixed luminosity of $\log
L/L_{\odot}$ = 5.3, were scaled to match the observation. The parameters $L$
and $T_{*}$ were used to derive the stellar radius from Eq.\,(\ref{eq:lrt}), 
and from the ``transformed radius'' $R_{\mathrm{t}}$ we obtained the mass-loss 
rate via Eq.\,(\ref{eq:rt}) under the assumption of $D$\,=\,10. The final 
values of $R_{*}$, $\dot{M}$, and $\log\,L$ are included in   
Table\,\ref{tab:wcpar}. 

\begin{figure*}[bth]
  \sidecaption
  \includegraphics[width=12cm]{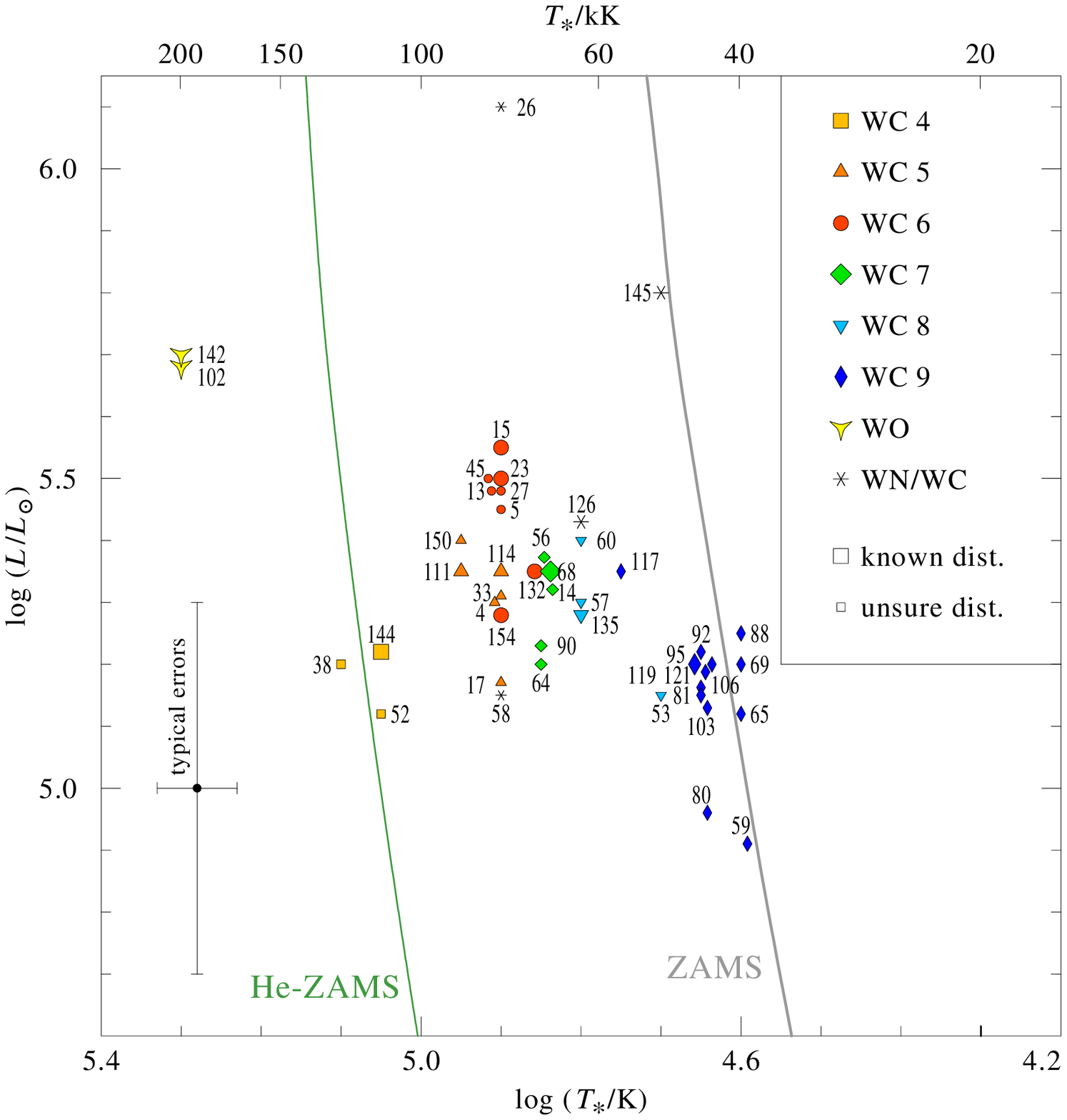}
  \caption{Hertzsprung-Russell diagram of the Galactic WC stars analyzed
    in this work. Stars with known distances are represented by larger symbols.
		The symbols are labeled with the WR catalog number
    \citep{vdH2001} of the corresponding stars.}
  \label{fig:hrd}
\end{figure*}  

  From the luminosities and the stellar temperature, we can construct
the empirical Hertzsprung-Russell diagram (HRD, Fig.\,\ref{fig:hrd}).
The stars with distances known from cluster or association memberships
are represented by larger symbols, because their luminosities are more
trustworthy than those relying on the subtype calibration.

  As one can see, the WC stars are located between the hydrogen and 
helium zero-age main-sequence (ZAMS), except for some of the WC stars 
that fall on the cooler side of the hydrogen ZAMS. The two WO stars 
have temperatures above those that would be expected from the He-ZAMS.
The WC subtypes form a sequence in the HRD, from the WC9 stars, which 
are the coolest stars with the lowest luminosities, to the WC4
stars, which are hotter and more luminous. It is unclear whether the
scatter is intrinsic to the stars, or caused by wrong distances. Taking the
luminosities at face value, they range from about $\log L/L_{\odot} = 
5.0$ for the faintest WC9 star to about $\log L/L_{\odot} = 5.8$ for the
brightest WC6 star and the two WOs. Two of the WN/WC transition type
stars (WR\,26 and WR\,145) have exceptionally high luminosities, which
are rather typical of WN stars. The HRD positions are discussed
further in Sect.\,\ref{sec:evolution} as regards stellar evolution.

\begin{figure}[bth]
  \resizebox{\hsize}{!}{\includegraphics{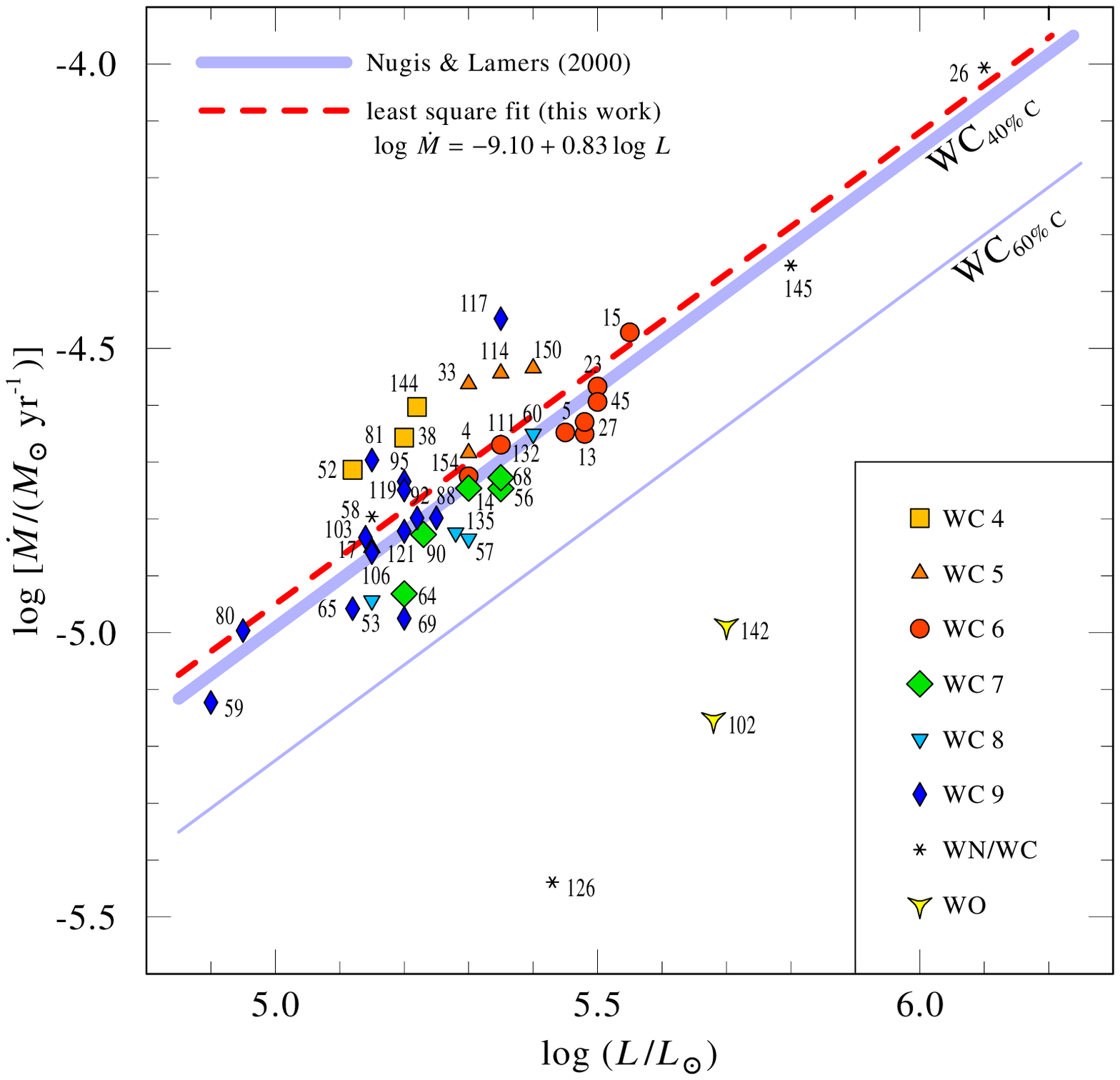}}
  \caption{Empirical mass-loss rates versus luminosity for the analyzed 
  Galactic WC stars. The subtypes are distinguished by different symbols 
  as explained in the inlet. The numbers beside the symbols identify 
  each star by its number in the WR catalog \citep{vdH2001}. The 
  red-dashed red line gives the least square fit to these results. The 
  full, shaded lines represent the relations proposed by \citet{NL2000mdot} 
  for WC stars with 40\% and 60\% carbon, respectively.}
  \label{fig:mdotl}
\end{figure}

  All our program stars show strong mass-loss. The empirical mass-loss 
rates from Table\,\ref{tab:wcpar} are plotted versus their corresponding
luminosities in Fig.\,\ref{fig:mdotl}, and follow a tight linear correlation 
in the double-logarithmic diagram, except for the WO stars. This is
unsurprising if we remember that the WC stars basically follow the proportionality
$R_{\mathrm{t}} \propto T_{*}^2$ (cf.\ Eq.\,\ref{eq:wctrend}). If all
stars had the same $\varv_\infty$, this would yield the mass
loss-luminosity relation $\dot{M} \propto L^{3/4}$. This slope with a
power of $3/4$ was already suggested by \citet{LMC1998}, who combined 
analyses from WC stars in the LMC and the Galaxy. The actual trend in
Fig.\,\ref{fig:mdotl} is slightly steeper, because $\varv_\infty$ also
increases towards the earlier subtypes. The least square fit to our data 
gives 
\begin{equation}
  \label{eq:mdotlfit}
  \log \frac{\dot{M}}{M_{\odot}/\mathrm{yr}} = \left( 0.83 \pm 0.11 \right) \log \frac{L}{L_{\odot}} - 9.10 \pm 0.57
\end{equation}
Our empirical data confirm amazingly well the relation 
suggested by \citet{NL2000mdot} -- cf. Eq.\,(21) therein -- based
on the empirical data available at that time, when we assume a 
heavy element mass fraction in their formula of 45\% (thick blue line in 
Figure\,\ref{fig:mdotl}). The value of 45\% results from our model 
assumptions of 40\% carbon and 5\% oxygen.

  Two more columns are included in Table\,\ref{tab:wcpar}. The current 
stellar mass $M$ is deduced from the luminosity. For WC stars, the 
corresponding relation from \citet{Langer89mass} was applied, while 
for the transition types we adopted the relation for hydrogen-free WN 
stars from the same reference.  
The last column in Table\,\ref{tab:wcpar} lists the values for the wind
efficiency parameter $\eta$, defined as the ratio of wind momentum
$\dot{M} \varv_{\infty}$ and photon momentum $L/c$ per time:
\begin{equation}
  \label{eq:eta}
  \eta := \frac{\dot{M}\,\varv_{\infty}\,c}{L}
\end{equation}
The ``single scattering limit'', at which every photon is scattered
exactly once to accelerate the wind, corresponds to $\eta = 1$. Hence,
each photon has to be scattered on average $\eta$ times to drive the
wind with the empirical parameters given in the table. \citet{GH05} 
constructed a hydrodynamically self-consistent model for the WC5 star 
WR\,111 that achieved $\eta$ = 2.54, thus showing that 
efficiency values of a few can be reached theoretically
when multiple-scattering effects are properly taken into
account. However, it seems questionable whether values of $\eta \goa 5$
can be explained. In our results many stars have wind efficiencies of 
$\eta > 5$, or even $\eta > 10$, especially among the early WC subtypes.
Interestingly, the two WO stars have only moderate $\eta$ values.  

Possible solutions to this wind-driving problem include: (a) multiple
scattering is even more efficient than hitherto thought, (b) there is an
additional, yet unidentified  wind-driving mechanism, and (c) our
mass-loss rates are drastic overestimates because the clumping
factor is much higher than the value of $D$=10 that we adopted.

\begin{table*}
  \caption{Parameters of the Galactic single WC stars}
  \label{tab:wcpar}
  \centering

  \begin{tabular}{r l r r l l rcr r r l r r}
      \hline
      \hline
\multicolumn{1}{c}{WR} & \multicolumn{1}{l}{Subtype} & \multicolumn{1}{c}{$T_{*}$} & \multicolumn{1}{c}{$\log R_{\mathrm{t}}$} & \multicolumn{1}{c}{$\varv_{\infty}$} & \multicolumn{1}{c}{$E_{b-\varv}$} & \multicolumn{1}{c}{D.M.} & \multicolumn{1}{c}{\rule[0mm]{0mm}{3mm}} & \multicolumn{1}{c}{$M_{\varv}$} & \multicolumn{1}{c}{$R_{*}$} & \multicolumn{1}{c}{$\log\dot{M}$} & \multicolumn{1}{c}{$\log L$} & \multicolumn{1}{c}{$M$\tablefootmark{a}} & \multicolumn{1}{c}{$\eta$} \\
& & \multicolumn{1}{c}{[kK]} & \multicolumn{1}{c}{[$R_{\odot}$]} & \multicolumn{1}{c}{[km/s]} & \multicolumn{1}{c}{[mag]} & \multicolumn{1}{c}{[mag]} & & \multicolumn{1}{c}{[mag]} & \multicolumn{1}{c}{[$R_{\odot}$]} & \multicolumn{1}{c}{[$M_{\odot}/\mathrm{yr}$]} & \multicolumn{1}{c}{[$L_{\odot}$]} & \multicolumn{1}{c}{[$M_{\odot}$]} & \\
      \hline
102 & WO2\rule[0mm]{0mm}{3mm} & 200 & 0.4 & 5000 & 1.08 & 12.39 & {\tiny$\rightarrow$} & $-$1.71 & 0.58 & $-$5.15 & 5.68 & 19 & 3.6  \\
142 & WO2 & 200 & 0.3 & 5000 & 1.43 & 10.45 & {\tiny$\rightarrow$} & $-$2.49 & 0.59 & $-$4.99 & 5.7 & 20 & 5.0  \\
38 & WC4\rule[0mm]{0mm}{5mm} & 126 & 0.1 & 3200 & 1.11 & 14.20 & {\tiny$\leftarrow$} & $-$3.34 & 0.84 & $-$4.66 & 5.2 & 10 & 21.9  \\
52 & WC4 & 112 & 0.2 & 3225\tablefootmark{b} & 0.56 & 11.35 & {\tiny$\leftarrow$} & $-$3.34 & 0.96 & $-$4.71 & 5.12 & 9 & 23.3  \\
144 & WC4 & 112 & 0.2 & 3500 & 1.6 & 11.3 & {\tiny$\rightarrow$} & $-$3.34 & 1.08 & $-$4.60 & 5.22 & 11 & 25.9  \\
4 & WC5\rule[0mm]{0mm}{5mm} & 79 & 0.5 & 2528\tablefootmark{c} & 0.6 & 11.83 & {\tiny$\leftarrow$} & $-$4.12 & 2.37 & $-$4.68 & 5.3 & 12 & 12.9  \\
17 & WC5 & 79 & 0.5 & 2231\tablefootmark{c} & 0.31 & 13.11 & {\tiny$\leftarrow$} & $-$4.12 & 1.99 & $-$4.85 & 5.15 & 10 & 11.0  \\
33 & WC5 & 79 & 0.5 & 3342\tablefootmark{c} & 0.6 & 14.07 & {\tiny$\leftarrow$} & $-$4.12 & 2.37 & $-$4.56 & 5.3 & 12 & 22.5  \\
111 & WC5 & 89 & 0.4 & 2398\tablefootmark{c} & 0.34 & 11.0 & {\tiny$\rightarrow$} & $-$4.16 & 1.99 & $-$4.67 & 5.35 & 12 & 11.3  \\
114 & WC5 & 79 & 0.5 & 3200 & 1.35 & 11.5 & {\tiny$\rightarrow$} & $-$4.08 & 2.51 & $-$4.54 & 5.35 & 12 & 20.1  \\
150 & WC5 & 89 & 0.4 & 3000 & 0.8 & 13.83 & {\tiny$\leftarrow$} & $-$4.12 & 2.11 & $-$4.53 & 5.4 & 13 & 17.1  \\
5 & WC6\rule[0mm]{0mm}{5mm} & 79 & 0.5 & 2120\tablefootmark{c} & 0.85 & 12.17 & {\tiny$\leftarrow$} & $-$4.42 & 2.81 & $-$4.65 & 5.45 & 14 & 8.3  \\
13 & WC6 & 79 & 0.5 & 2000 & 1.21 & 13.24 & {\tiny$\leftarrow$} & $-$4.42 & 2.91 & $-$4.65 & 5.48 & 15 & 7.3  \\
15 & WC6 & 79 & 0.5 & 2675\tablefootmark{c} & 1.23 & 11.28 & {\tiny$\rightarrow$} & $-$4.60 & 3.16 & $-$4.47 & 5.55 & 16 & 12.5  \\
23 & WC6 & 79 & 0.5 & 2342\tablefootmark{c} & 0.55 & 11.8 & {\tiny$\rightarrow$} & $-$4.64 & 2.98 & $-$4.57 & 5.5 & 15 & 9.9  \\
27 & WC6 & 79 & 0.5 & 2100 & 1.4 & 12.63 & {\tiny$\leftarrow$} & $-$4.42 & 2.91 & $-$4.63 & 5.48 & 15 & 8.0  \\
45 & WC6 & 79 & 0.5 & 2200 & 1.44 & 13.32 & {\tiny$\leftarrow$} & $-$4.42 & 2.98 & $-$4.59 & 5.5 & 15 & 8.7  \\
132 & WC6 & 71 & 0.6 & 2400 & 1.15 & 13.16 & {\tiny$\rightarrow$} & $-$4.38 & 3.15 & $-$4.67 & 5.35 & 12 & 11.3  \\
154 & WC6 & 79 & 0.5 & 2300 & 0.78 & 12.2 & {\tiny$\rightarrow$} & $-$4.05 & 2.37 & $-$4.72 & 5.3 & 12 & 10.7  \\
14 & WC7\rule[0mm]{0mm}{5mm} & 71 & 0.6 & 2194\tablefootmark{c} & 0.65 & 10.55 & {\tiny$\leftarrow$} & $-$4.18 & 2.98 & $-$4.75 & 5.3 & 12 & 9.7  \\
56 & WC7 & 71 & 0.6 & 2009\tablefootmark{c} & 0.7 & 15.25 & {\tiny$\leftarrow$} & $-$4.18 & 3.15 & $-$4.75 & 5.35 & 12 & 7.9  \\
64 & WC7 & 71 & 0.6 & 1700 & 1.2 & 14.85 & {\tiny$\leftarrow$} & $-$4.18 & 2.65 & $-$4.93 & 5.2 & 10 & 6.2  \\
68 & WC7 & 71 & 0.6 & 2100 & 1.4 & 12.57 & {\tiny$\rightarrow$} & $-$4.18 & 3.15 & $-$4.73 & 5.35 & 12 & 8.6  \\
90 & WC7 & 71 & 0.6 & 2053\tablefootmark{c} & 0.4 & 9.50 & {\tiny$\leftarrow$} & $-$4.18 & 2.75 & $-$4.83 & 5.23 & 11 & 8.8  \\
53 & WC8d\rule[0mm]{0mm}{5mm} & 50 & 0.9 & 1800 & 0.75 & 12.29 & {\tiny$\leftarrow$} & $-$4.48 & 5.00 & $-$4.94 & 5.15 & 10 & 7.1  \\
57 & WC8 & 63 & 0.7 & 1787\tablefootmark{c} & 0.38 & 12.72 & {\tiny$\leftarrow$} & $-$4.48 & 3.75 & $-$4.84 & 5.3 & 12 & 6.4  \\
60 & WC8 & 63 & 0.7 & 2300 & 1.45 & 11.79 & {\tiny$\leftarrow$} & $-$4.48 & 4.21 & $-$4.65 & 5.4 & 13 & 10.1  \\
135 & WC8 & 63 & 0.6 & 1343\tablefootmark{c} & 0.4 & 11.2 & {\tiny$\rightarrow$} & $-$4.48 & 3.66 & $-$4.82 & 5.28 & 11 & 5.2  \\
59 & WC9d\rule[0mm]{0mm}{5mm} & 40 & 1.0 & 1300 & 2.0 & 10.73 & {\tiny$\leftarrow$} & $-$5.17 & 5.94 & $-$5.12 & 4.9 & 7 & 6.1  \\
65 & WC9d & 40 & 1.0 & 1300 & 2.0 & 10.73 & {\tiny$\leftarrow$} & $-$5.17 & 7.66 & $-$4.96 & 5.12 & 9 & 5.3  \\
69 & WC9d & 40 & 1.0 & 1089\tablefootmark{c} & 0.55 & 12.39 & {\tiny$\leftarrow$} & $-$5.17 & 8.40 & $-$4.97 & 5.2 & 10 & 3.6  \\
80 & WC9d & 45 & 0.9 & 1600 & 1.8 & 12.20 & {\tiny$\leftarrow$} & $-$5.17 & 4.99 & $-$5.00 & 4.95 & 8 & 8.9  \\
81 & WC9 & 45 & 0.8 & 1600 & 1.5 & 11.40 & {\tiny$\leftarrow$} & $-$5.17 & 6.28 & $-$4.70 & 5.15 & 10 & 11.2  \\
88 & WC9 & 40 & 1.0 & 1500 & 1.4 & 12.13 & {\tiny$\leftarrow$} & $-$5.17 & 8.89 & $-$4.80 & 5.25 & 11 & 6.6  \\
92 & WC9 & 45 & 0.8 & 1121\tablefootmark{c} & 0.52 & 13.60 & {\tiny$\leftarrow$} & $-$5.17 & 6.81 & $-$4.80 & 5.22 & 11 & 5.3  \\
95 & WC9d & 45 & 0.9 & 1900 & 1.74 & 11.61 & {\tiny$\rightarrow$} & $-$5.17 & 6.66 & $-$4.73 & 5.2 & 10 & 10.9  \\
103 & WC9d & 45 & 0.8 & 1190\tablefootmark{b} & 0.52 & 11.90 & {\tiny$\leftarrow$} & $-$5.17 & 6.21 & $-$4.83 & 5.14 & 10 & 6.2  \\
106 & WC9d & 45 & 0.8 & 1100 & 1.2 & 12.43 & {\tiny$\leftarrow$} & $-$5.17 & 6.28 & $-$4.86 & 5.15 & 10 & 5.3  \\
117 & WC9d & 56 & 0.6 & 2000 & 1.56 & 12.95 & {\tiny$\leftarrow$} & $-$5.17 & 5.00 & $-$4.45 & 5.35 & 12 & 15.7  \\
119 & WC9d & 45 & 0.8 & 1300 & 0.9 & 13.98 & {\tiny$\leftarrow$} & $-$5.17 & 6.66 & $-$4.75 & 5.2 & 10 & 7.2  \\
121 & WC9d & 45 & 0.8 & 1100 & 1.40 & 11.84 & {\tiny$\leftarrow$} & $-$5.17 & 6.66 & $-$4.82 & 5.2 & 10 & 5.1  \\
26 & WN7/WCE\rule[0mm]{0mm}{5mm} & 79 & 0.6 & 2700 & 1.25 & 15.19 & {\tiny$\leftarrow$} & $-$5.67 & 5.95 & $-$4.01 & 6.1 & 37 & 10.4  \\
58 & WN4/WCE & 79 & 0.5 & 1600 & 0.55 & 14.50 & {\tiny$\leftarrow$} & $-$3.84 & 1.99 & $-$4.80 & 5.15 & 10 & 8.9  \\
126 & WC5/WN & 63 & 1.2 & 2000 & 0.95 & 13.22 & {\tiny$\rightarrow$} & $-$3.82 & 4.35 & $-$5.44 & 5.43 & 14 & 1.3  \\
145 & WN7/WCE & 50 & 0.9 & 1440 & 1.86 & 11.3 & {\tiny$\rightarrow$} & $-$6.38 & 10.57 & $-$4.35 & 5.8 & 25 & 5.0  \\
    \hline
  \end{tabular}
  \tablefoot{
     \tablefoottext{a}{Masses are calculated from luminosity after \citet{Langer89mass} using his WC relation for WC stars and his WNE relation for WN/WC stars.}
     \tablefoottext{b}{\citet{Prinja90}}
     \tablefoottext{c}{\citet{NiSk2002}}\\
     The arrows indicate whether we infer the absolute magnitude
       ($M_{\varv}$) from a known distance modulus D.M., or whether 
       the distance is calculated from $M_{\varv}$ as obtained by the 
       subtype magnitude calibration (cf.\,Table\,\ref{tab:wcmvcalib}).
  }
\end{table*}

An individual and detailed determination of the \emph{chemical
composition} of our program stars is beyond the scope of the present
paper. For the WC stars, we checked that the composition adopted for our
model grid (with 40\% carbon, cf.\,Table\,\ref{tab:grids}) is roughly
adequate. This can best be seen from the neighboring ``diagnostic line
pair'' \ion{He}{ii}\,5412\,\AA\ and \ion{C}{iv}\,5470\,\AA. These two
lines form in the same zone of the wind and display a very similar
dependence on $T_{*}$ and $R_{\mathrm{t}}$, while the ratio of their
line strengths depends sensitively on the C/He abundance ratio. 
We find that a mass fraction ratio C:He of 40:55 is adequate for all 
WC stars in our sample, including the WC9 subtypes. As suggested 
by \citet{Crowther2006}, there is no indication that WC9 stars are 
chemically less evolved than the earlier subtypes.

  Overall, models with an oxygen mass fraction of 5\% seem to be appropriate
for WC stars. However, the situation for the individual oxygen-line fit differs. 
The \ion{O}{iv}-line at 3411\,\AA\ is usually nicely reproduced, while the fit quality
of the \ion{O}{v} line at 5590\,\AA, which is used for classification, varies even 
within the subtypes. For the WC4 stars WR\,52 and WR\,144, the line is significantly
stronger. Thus, we used a model with an enhanced oxygen fraction of 15\%.
This raises the question of whether there might be an oxygen trend throughout the WC
subtype sequence. There are indeed some stars of the subtypes WC5 and WC6 
where we see that \ion{O}{v}\,5590\,\AA\ is significantly stronger than 
predicted by the model, but there are also several examples where this line
is perfectly reproduced with no more than the standard 5\% oxygen mass fraction. We
therefore conclude that there is no general increase in oxygen throughout the 
WC sequence, although there may possibly be an increate for the WC4 subtype.

  The WO stars, however, require models with even more oxygen, roughly 
30\% (by mass) instead of the 5\% chosen for the WC grid. Interestingly 
the carbon abundance does not increase but stays at 40\%. The remaining 
approximately 30\% of the mass consists of helium, as the fraction of iron group
elements is also the same as in the WC models, namely 0.16\%.

  The spectra of the three WN/WC transition-type stars in our sample could 
not be fitted with models of WC-type composition. Their spectral appearance 
resembles more those of the WN than the WC stars. To simulate 
these spectra, we started from our models of hydrogen-free WN stars
\citep[see][]{Adriane06}. These models contain mostly (98\%) helium,
plus 1.5\% N and only 0.01\% C. While this nitrogen abundance also complies
with the N features in our WN/WC spectra, the carbon abundance has
to be increased in order to reproduce the C lines.

  We found that the adequate carbon abundances differ from star to star in this 
very special group. In particular, WR\,58 (WN4/WCE) and WR\,145 (WN7/WCE)
contain a small amount of carbon, namely 0.1\% and 0.5\% by mass,
respectively. In contrast, WR\,26 (WN7/WCE) even requires models with a 
carbon mass fraction as high as 20\% to reproduce its spectrum. We note 
that we have omitted oxygen in our WN/WC special models to accelerate our
calculations, which seems justified because the observed WN/WC spectra 
show hardly any oxygen features.

  The WC/WN star WR\,126 is unique to our sample. The only notable
emission lines are the C-He-blend around 4650\,\AA, the
\ion{C}{iv}\,5808\AA-line, and three minor lines. Two of them can easily
be associated with \ion{He}{ii}. The peak around 7100\,\AA, however,
does not match any line usually seen in WC spectra. However, WN
stars display a blend of \ion{N}{iv} lines in this place. For this reason,
this star is classified as WC5/WN. For our fit, we used a nitrogen-free
WC star model with 20\% (i.e.\ reduced) carbon mass fraction and 5\% oxygen.

  Hence, it seems that for the transition-type stars neither the carbon
abundance nor the position in the $\log R_{\mathrm{t}}$-$\log
T_{*}$-plane (see Fig.\,\ref{fig:wcstats}) correlates with the spectral
subtype.

  Summarizing the results for our whole sample, we found that all 
parameters show a close correlation with the spectral subtype (except 
for the transition types to WN). In Table\,\ref{tab:wcmean}, we compile 
the mean parameters for each subtype. These results are discussed 
further in Sect.\,\ref{sec:evolution} as regards the stellar evolution.

\subsection{Error estimations}
  \label{subsec:errors}

  As explained in Sect.\,\ref{subsec:param}, the spectroscopic parameters 
($T_{\ast}$, $R_{\mathrm{t}}$) were determined by selecting the best-fitting 
synthetic line spectrum from our model grids. Those grids are spaced by 0.05 
in $\log T_{\ast}$ and $0.1\,$dex in $R_{\mathrm{t}}$, which corresponds to 
our experience that this is roughly the accuracy with which the optimum line 
fit can be identified.
 
For very dense winds ($\log R_{\mathrm{t}}/R_{\odot} \la 0.4$), all emergent
radiation originates from the fast-moving parts of the stellar wind, as discussed
in Sect.\,\ref{subsec:param}. Hence, in this regime of parameter degeneracy
only the product $R_{\mathrm{t}}T_{\ast}^2$ is constrained by the line fit.

  While the spectroscopic parameters are derived from the (normalized)
line spectrum, the luminosity is obtained by fitting the SED. For most of our
program stars, the photometrically calibrated observations cover a wide spectral
range from the UV to the IR, hence the color excess $E_{b-\varv}$ can be
determined to an accuracy of $\pm$0.02\,mag. The uncertainty in the distance is
larger, which enters the derived luminosity $\propto d^2$ and the mass-loss
rate $\propto d^{3/2}$. For many of our stars we must rely on the assumption that
the $\varv$ magnitude is constant per subtype (Sect.\,\ref{sec:method}). This
whole calibration, in turn, is based on those stars that presumably belong to a
certain open cluster or association. Apart from the possibility that the membership
assignment may be erroneous in individual cases, the distance to the clusters or
associations are also continue to be debated between different authors, as particularly 
mentioned in appendix Sect.\,\ref{appsec:starcomments}. To our impression, 
the distance moduli remain uncertain by about 0.75\,mag, leading to an error 
margin of $\pm$0.3\,dex for the luminosity.

  Systematic errors of our analyses are extremely difficult to quantify.
Atomic data are one source of uncertainties. A more important question is the degree
to the basic model assumptions, such as spherical symmetry and homogeneity,
are adequate for real stars. A plethora of recent work deals with \emph{clumping in hot
star winds} \citep[cf.][]{ClumpingWS2008}. Clumping can bias in particular the empirical
mass-loss rates.

\subsection{Comparison with previous WC analyses}
  \label{subsec:compare}

A larger sample of Galactic WC stars were analyzed in our group by
\citet{KH95}, yet with un-blanketed PoWR model atmospheres. Their results 
did not reveal the clear correlation between the subtypes and the stellar
parameters that we see now. The former analysis was partly confused by a couple
of objects that have since been identified as binaries. In
the case of WN stars, the improvement from un-blanketed to
iron-line-blanketed models led to about 10\,kK higher stellar
temperatures \citep{Adriane06}. Surprisingly, this is not the case for the
WC stars. The newly obtained mass-loss rates are lower, mainly because of
the microclumping correction (cf.\ Sect.\,\ref{sec:grid}). 

  A couple of Galactic WC stars were analyzed more recently with 
the help of the line-blanketed CMFGEN model atmosphere code by
\citet{HiMi98}. WR\,111 (WC5) was analyzed by \citet{HiMi99}.
\citet{Dessart2000} studied another four Galactic WC stars, two of them,
WR\,90 (WC7) and WR\,135 (WC8), being single stars.
\citet{Crowther2006} analyzed the WC9d star WR\,103. Their results are
compared in Table\,\ref{tab:wcparcmp} with those obtained in this work
(``t.w.''). The agreement among the stellar temperatures is very good. We
tend to obtain slightly higher luminosities by 0.1\,dex. The mass-loss
rates have a considerable scatter; we note that they were derived with the 
same assumption for the clumping parameter ($D = 10$) as in this work.

  Interestingly, \citet{Dessart2000} also concluded that there is no
correlation between stellar temperature and subtype, as in our 
\citet{KH95} paper, and in sharp contrast to our present result. 
The reason is obviously that their study was also confused by 
binaries that have diluted emission lines.

\begin{table}
  \caption{Comparison of stellar parameters with previous analyses}
    
  \label{tab:wcparcmp}
  \centering

  \begin{tabular}{r c c r l l l}
     \hline 
     \hline 
        WR & Subtype &    Ref.    & \multicolumn{1}{c}{$T_{*}$}
                                            & \multicolumn{1}{c}{$\log\,L$}
                                                           & \multicolumn{1}{c}{$\varv_{\infty}$} 
                                                                                      & $\log\,\dot{M}$ \rule[0mm]{0mm}{3mm} \\
           &         &            & \multicolumn{1}{c}{[kK]}   
                                            & \multicolumn{1}{c}{[$L_{\odot}$]} 
                                                           & \multicolumn{1}{c}{[km/s]}      
                                                                                      & [$M_{\odot}$/yr] \\
     \hline 
      111  &   WC5   &    t.w.    &   89    &     5.35     &  2398\tablefootmark{(4)} &   $-$4.67 \rule[0mm]{0mm}{3mm}  \\
           &         &    (1)     &   91    &     5.3      &  2300                    &   $-$4.8    \\
           
       90  &   WC7   &    t.w.    &   71    &     5.23     &  2053\tablefootmark{(4)} &   $-$4.83   \\
           &         &    (2)     &   71    &     5.5      &  2045                    &   $-$4.6    \\
      135  &   WC8   &    t.w.    &   63    &     5.28     &  1343\tablefootmark{(4)} &   $-$4.82   \\
           &         &    (2)     &   63    &     5.2      &  1400                    &   $-$4.9    \\
      103  &   WC9d  &    t.w.    &   45    &     5.14     &  1190\tablefootmark{(5)} &   $-$4.83   \\
           &         &    (3)     &   48    &     4.90     &  1140                    &   $-$4.50   \\
     \hline 
  \end{tabular}

  \tablebib{
  		 (1)~\citet{HiMi99}; (2)~\citet{Dessart2000}; (3)~\citet{Crowther2006}; 
  		 (4)~\citet{NiSk2002}; (5)~\citet{Prinja90}
  }

\end{table}
  
\section{Stellar evolution}
  \label{sec:evolution}

\begin{figure}[bth]
  \resizebox{\hsize}{!}{\includegraphics{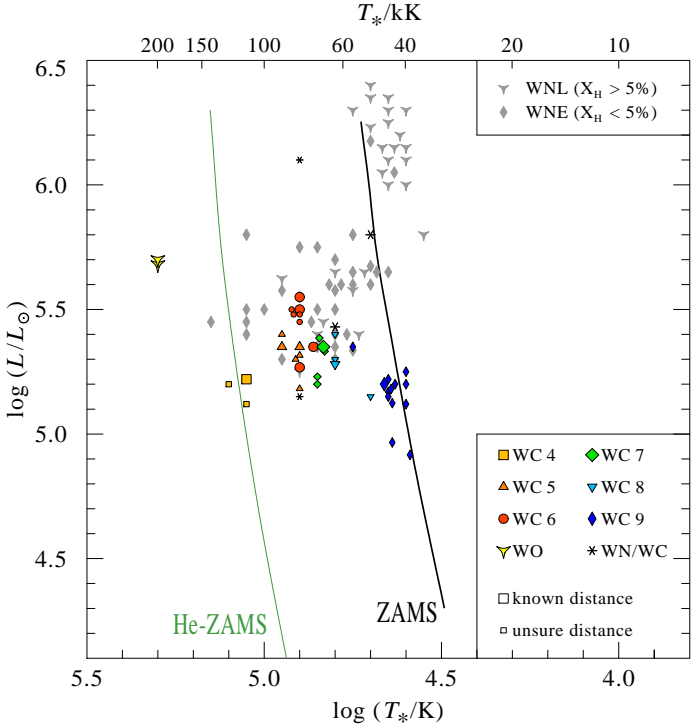}}
  \caption{HRD with the WC star positions from this work and the
    Galactic WN star positions from \citet{Adriane06}.}
  \label{fig:hrd_wcpluswn}
\end{figure}

  To discuss the evolutionary origin of the WC stars, we now compare
their empirical positions in the HRD with those of their potential
progenitors, the WN stars. The results of this work are shown in
Fig.\,\ref{fig:hrd_wcpluswn}, together with those of the Galactic WN stars 
studied by \citet{Adriane06}. The WN stars form two distinct groups, the very
luminous ``late'' subtypes (WNL) with a significant fraction of hydrogen
in their atmosphere, and the less-luminous ``early'' subtypes (WNE),
most of which are completely hydrogen-free.

  Like the WNE stars, the WC stars group at the left, hotter side of the ZAMS,
while the hydrogen-containing WNL stars lie on the cooler side of the
main sequence. The WO stars are even hotter than stars on the helium zero-age
main sequence would be. We note that the luminosity range for the WC stars of
early subtypes (WC4--7) and two WO stars is almost the same as for the hydrogen-free 
WNE subclass. The latest WC subtypes, WC8--9, are less luminous than any WN
star. The large group of WNL stars, which are very luminous 
($\log L/L_\odot \geq 6$), have no counterparts to similarly luminous WC stars.

\begin{figure}[bth]
  \resizebox{\hsize}{!}{\includegraphics{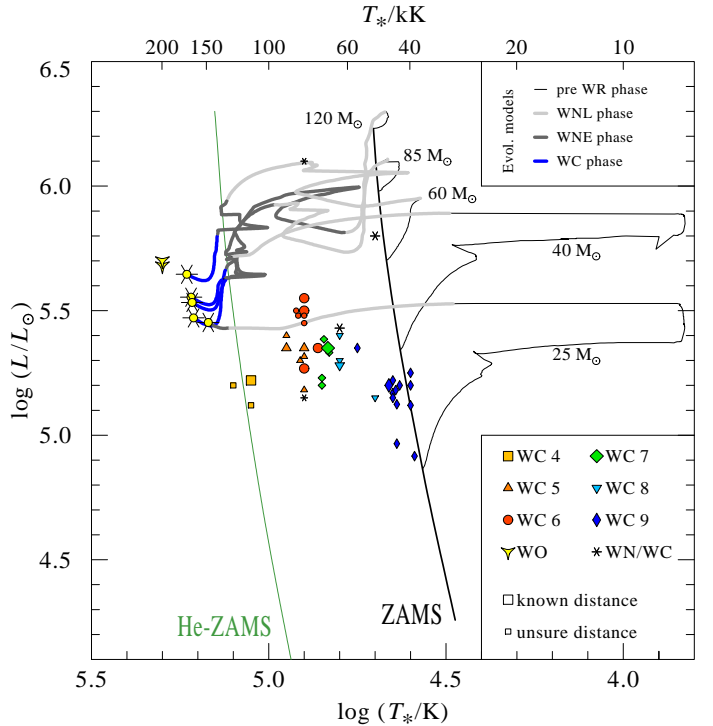}}
  \caption{HRD with the WC star positions and the tracks from
    \citet{Geneva2003} with an initial rotational velocity of
    $\varv_{\mathrm{rot}}\,=\,300\,$km/s. The thick lines indicate 
    the WR phases of the tracks.}
  \label{fig:hrd_geneva_rot}
\end{figure}

\begin{figure}[bth]
  \resizebox{\hsize}{!}{\includegraphics{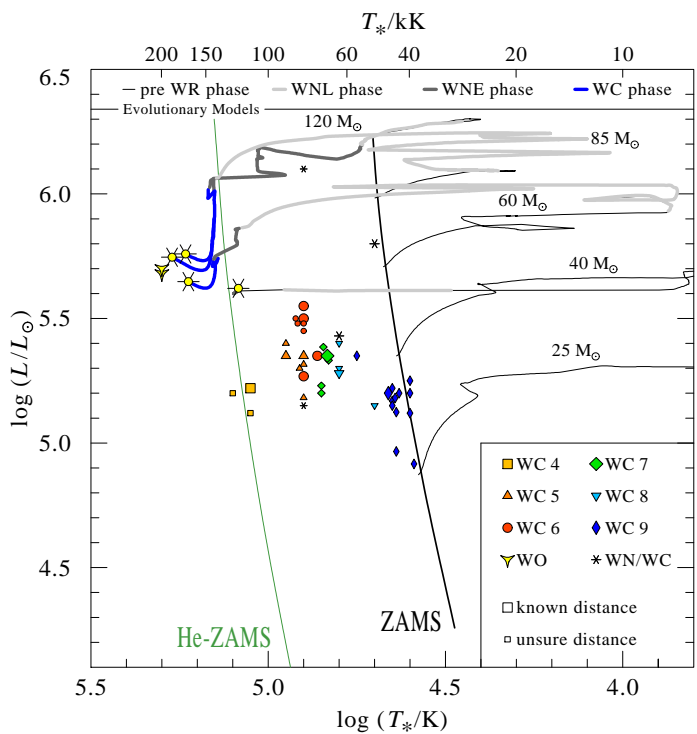}}
  \caption{HRD with the WC star positions and the tracks from 
    \citet{Geneva2003} without rotation. The thick lines indicate 
    the WR phases of the tracks.}
  \label{fig:hrd_geneva_norot}
\end{figure}

  These results provide an interesting test case for stellar evolution models. 
The stellar evolution tracks calculated by \citet{Geneva2003} from the Geneva group 
are the standard reference for massive-star evolution. In Figs.\,\ref{fig:hrd_geneva_rot} 
and \ref{fig:hrd_geneva_norot}, we compare the HRD of our WC sample with the
tracks of \citet{Geneva2003} in the different versions with and
without accounting for rotation, respectively.  According to the
chemical composition at the surface, the tracks are drawn in different
styles: hydrogen-rich phases are represented by thin black lines.  When
the hydrogen mass fraction of the atmosphere drops below 40\% as in the 
WNL stars, the track is shown as a thick gray line. The hydrogen-free WN
stage ($X_{\rm H} < 5\%$) is indicated by a thick dark line. In the WC stage 
(thick black/blue line), carbon reaches more than 20\% in the stellar envelope. 
We chose this limit to be higher than in \citet{Adriane06} to ensure 
that the WC part of the tracks exclude any WN/WC transition stage.

  The evolutionary tracks predict that in the non-rotating scenario, massive stars 
with an initial mass of 60\,$M_{\odot}$ and above undergo a luminous blue variable (LBV) 
phase before entering the WNL stage. However, in the scenarios with rotation these 
stars skip the LBV phase and enter the WNL stage directly after the main
sequence \citep{Geneva2003}. Afterwards, the WNL stars evolve to 
hydrogen-free WNE stars at slightly lower luminosities, depending on the initial mass. 
They eventually enter the WC stage at roughly the helium main-sequence. According to 
the continuous mass-loss and the mass-luminosity relation, their luminosity decreases 
by a few tenths in $\log L$ before the gravitational core-collapse.

{Comparing now the Geneva tracks with our empirical HRD
(Figs.\,\ref{fig:hrd_geneva_rot} and \ref{fig:hrd_geneva_norot}),
we find relatively good agreement for the two WO stars of our sample.
In particular the tracks without rotation and for initial masses 
$\ge 60\,M_{\odot}$ reach very hot and luminous endpoints prior to the 
supernova (SN) explosion.

For the WC stars, however, the comparison ends less favorably. The
first problem is posed by the luminosities. The evolutionary calculations
without rotation (Fig.\,\ref{fig:hrd_geneva_norot}) predict that only tracks
for initial masses above 37\,$M_{\odot}$ return from the red supergiant (RSG)
stage to the blue side of the HRD. The implied luminosities are higher than
those inferred from our WC sample.
The tracks with rotation (Fig.\,\ref{fig:hrd_geneva_rot}) predict WR stars
for initial masses above 22\,$M_\odot$. Hence, the tracks are at least
compatible with the most luminous WC stars of our sample, the WC6 subtypes.
The bulk of WC stars, however, are less luminous than predicted by any of the
tracks in their post-RSG phase. If we apply these relations between mass and
luminosity from the Geneva tracks, the conclusion is that WC stars mainly
arise from initial masses lower than 40\,$M_{\odot}$.}

  Our comparison of effective temperatures has some \emph{caveats}, since these
depend on the choice of the reference radius. The evolutionary tracks are
plotted over the stellar effective temperature that refers to the
hydrostatic stellar core. Our stellar temperature $T_{\ast}$ refers to the
radius of Rosseland optical depth 20 (cf.\ Sect.\,\ref{sec:grid}).  For
the adopted velocity (and implied density) structure, this is close to
the hydrostatic radius for most of the best-fit models. However, there are
indications \citep{Inflation1999, Inflation2006} that WR stars may have 
very extended, sub-photospheric layers, i.e.\ the hydrostatic core may 
actually be much smaller and hotter than derived from the observation of 
the stellar atmosphere. Moreover, for stars in the domain of the parameter
degeneracy (i.e. WC4--5) discussed in Sect.\,\ref{sec:results}, the hydrostatic
stellar radius can only be estimated from a significant inward
extrapolation of the velocity field into optically thick layers, which
can lead to either an over- or under-estimation of the corresponding 
effective temperature.  

  The empirical HRD of the WC stars shows a temperature gap between the
helium main-sequence and the WC positions. While the WO
subtypes are located to the left of the He-ZAMS (presumably indicating
the position of a hypothetical CO-ZAMS), the later WC subtypes are
progressively cooler. The later the WC subtype, the lower the
luminosity (implying a lower current mass according to the
mass-luminosity relation), and the larger must be the sub-photospheric
layer enhancing the stellar radius on top of the hydrostatic core.

\begin{figure}[bth]
  \resizebox{\hsize}{!}{\includegraphics{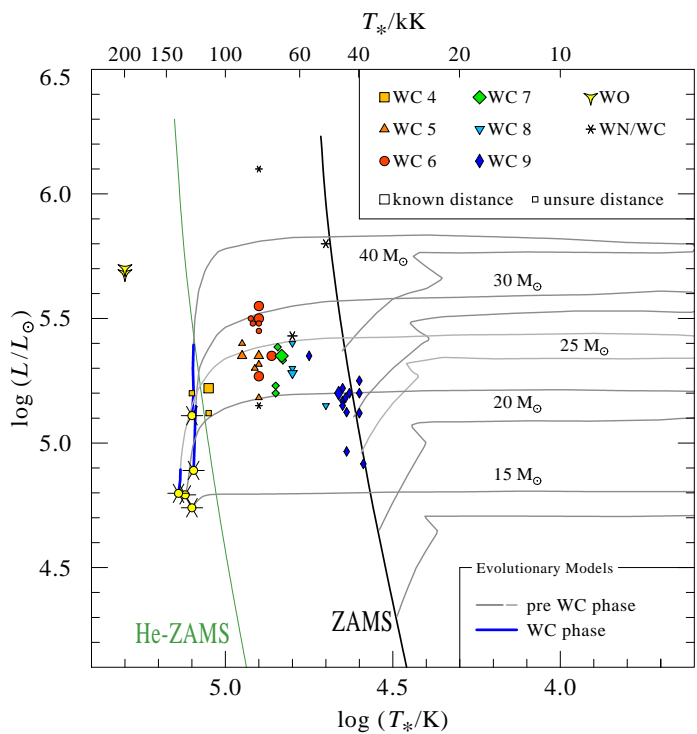}}
  \caption{HRD with the WC star positions and the tracks (thick lines)
    from \citet{Vanbeveren98} using higher mass-loss rates during the 
    RSG phase. {The track for $M_{\mathrm{init}}\,=\,25\,M_{\odot}$ 
    is shown in light grey for clarity. In the WC phase, all tracks are 
    drawn as thick dark/blue lines.}}
  \label{fig:hrd_vanbeveren}
\end{figure}

  One could argue that all luminosity values are affected by (almost)
all distance values for Galactic WR stars being inferred from cluster memberships. Even 
if some of the distances for individual stars were wrong due to a false assignment
to an association or cluster, only their HRD positions would change while
the overall picture would not be drastically affected, as long as cluster distances 
are not systematically underestimated. 

  In contrast to the Geneva stellar evolution models, \citet{Vanbeveren98}
showed that it is indeed possible to reach the WR stage for stars with a
minimum initial mass of only  15\,$M_{\odot}$, if the mass-loss rate
during the red super-giant (RSG) stage is higher than given by the
formula of \citet{deJager88} adopted for the Geneva
tracks. When adopting a higher RSG mass-loss, the resulting
tracks (see Fig.\,\ref{fig:hrd_vanbeveren}) predict, for instance, a
luminosity of $\log L/L_{\odot}\,\approx\,5.2$ for a star with
$M_{\mathrm{init}}\,=20\,M_{\odot}$ entering the post-RSG stage, 
and a final luminosity of $\log L/L_{\odot}\,\approx\,4.8$
before explosion. Although in the
models of \citet{Vanbeveren98}, WC phases only occur for stars with
$M_{\mathrm{init}}\,\geq\,25\,M_{\odot}$, the tracks cover almost the whole 
luminosity range of our WC sample. As for the results from the Geneva 
tracks, there is a discrepancy in the temperature $T_{\ast}$ between the
tracks and the evolutionary models. To date, none of the evolutionary 
calculations performed have been able to fully explain the obtained HRD 
positions, as they fail to reproduce the temperatures but provide
enough indications that WC stars originate from a lower mass range.
Both the Geneva and the Vanbeveren models imply that the WC phase is close
to the He-ZAMS. \citet{Inflation2011} showed that
extended sub-photospheric layers are not limited to hydrogen-rich stars on
the upper part of the main sequence, but also occur for H-free stars on the 
He-ZAMS. This happens even in a lower mass regime if the clumping factor 
is sufficiently high. As briefly discussed above, such an envelope 
inflation would strongly reduce the effective temperatures of the stars, 
placing them exactly in the temperature regime we obtained in our work. 
Inflation might be a key factor in solving the temperature discrepancy.

  Another crucial aspect is the highly uncertain mass-loss rates 
during the RSG stage pointed out by \citet{Meynet2011}, which is partly  
caused by dust. For objects covered with dust, \citet{vanLoon2005} 
discovered mass-loss rates that are significantly higher than those of visually 
bright ones. Such high rates would lead to stars evolving back 
from the red supergiant stage to the blue side of the HRD instead of 
exploding as a type II-P supernova. This would be in line with the observational 
constraints for type II-P supernovae provided by \citet{Smartt2009}, who found 
several SN II-P progenitors to be red supergiants, but none to have an 
initial mass of 18\,$M_{\odot}$ or higher. However, as there are known RSGs at 
least in the mass range between 18\,$M_{\odot}$ and 25\,$M_{\odot}$, it is likely 
that they do not explode, but instead evolve to blue supergiants and eventually 
WR stars. Blue supergiants that have previously undergone an RSG stage would be 
He-enriched. Objects with such an appearance have indeed been found 
by \citet{Przybilla2010}.

  \citet{Vanbeveren2007} also compared the parameters of the Galactic WN
stars from our previous study \citep{Adriane06} with his tracks, which 
incorporate a higher RSG mass-loss rate (but no rotationally induced
mixing), and concluded that their own tracks are in much better agreement with
our results than the Geneva tracks. 
   
  Tracks that lead to WC stars with lower initial masses question the
evolutionary scenarios of \citet{Conti79} and also the revised version
of \citet{LangerSzenario} that WNL stars evolve to WNE stars. 
The clear separation of WNL and WNE stars in the HRD found by
\citet{Adriane06} suggests instead that WNL stars may not manage to get rid of
their hydrogen layers, but explode as type II supernovae either directly
after their WNL phase or, more likely, after excursions into the LBV
domain. More recent results \citep[e.g.][]{Adriane2010, OfWNClassf} indeed
often discuss or assume that hydrogen-rich WNL (also labeled WNh) 
stars are core-hydrogen burning and cover the high-mass end of the main sequence. 
Simply put, with increasing initial mass, the spectral types at the high-mass
end of the main sequence change from Of to Of/WN and finally WNL.

  Moreover, our conclusions as regards stellar evolution can be explained
in consistence with the interpretation of supernovae statistics by
\citet{SNStats}. They analyzed the different fractions of core-collapse 
supernovae (CCSN) using the data from the Lick Observatory Supernovae
Search (LOSS), covering mostly host galaxies with metallicities similar to 
Galactic metallicities (0.5-2\,$Z_{\odot}$). If the WNL stars do explode as 
supernovae without losing their hydrogen envelope, they will produce a type II
supernova instead of the Ib and Ic supernovae that are usually considered to
be the end of WR evolution, because there is no hydrogen at all left in the 
WC stars and little in the WNE stars. \citet{SNSmith2008} discussed how type IIn
supernovae require stars with $M_{\mathrm{init}} > 50\,M_{\odot}$ and
these high mass stars can indeed explode without losing their
hydrogen envelope. As type IIn supernovae require huge eruptive mass-loss that
exceeds even the high WR mass-loss rates, the primary candidates are
LBVs, as suggested by \citet{LBVSN} and eventually proven for SN\,2005gl
\citep{LBVSNex}. The HRD positions obtained for 
the Galactic WNL stars \citep{Adriane06} corroborate our findings that the 
hydrogen-rich WNL stars are LBV progenitors which has also been suggested by 
\citet{WNLLBV}. All of the analyzed Galactic WNL stars seem to have initial 
masses higher than $60\,M_{\odot}$. Their evolutionary sequence
in the high mass range should therefore be  
  \begin{equation}
    \label{eq:evol60plus}
    \mathrm{Of} \rightarrow \mathrm{Of/WNL} \leftrightarrow \mathrm{LBV} 
           \left[ \rightarrow \mathrm{WNL}_{\mathrm{H-poor}} \right] \rightarrow \mathrm{SN\,IIn}. 
  \end{equation}
    
  There is an ongoing controversial debate about whether LBVs can be direct SN 
progenitors. In general, evolutionary calculations do not predict such a direct 
connection. The LBVs are characterized by their variability, which one cannot directly 
measure after the star has exploded, as e.g.\, in the case of SN2005gl \citep{LBVSNex}.
\citet{LBVSNSuggestion} suggested LBVs as progenitors of certain supernovae due to 
signatures of progenitor variations in the SN spectrum. For SN 2005gj, \citet{LBVSNother} 
argued that line profiles similar to the observed ones could only be found in LBVs. 
Finally, luminosity and color, if these data exist, may indicate whether a SN progenitor 
was a blue star beyond the Humphrey-Davidson limit.

Since we know WNL stars with very different hydrogen abundances 
\citep{Adriane06,Martins2008,Adriane2010}, it is likely that there are multiple
excursions between a quiet WNL stage and LBV eruptions, as indicated by the
double-side arrow in Eq.\,(\ref{eq:evol60plus}).
  
  {\citet{Dwarkadas2011} questioned the scenario of LBVs being direct 
SN progenitors. He argued that the deduced mass-loss rates are usually based
on the assumption of $r^{-2}$-density stratifications, which might not be correct.
He further argues that there should be at least a short WR phase between an LBV 
outburst and a SN explosion. For a type IIn supernova, this phase would correspond 
to a WNL type with a low hydrogen abundance. We therefore add the suffix
``H-poor" to this possible post-LBV WNL stage in Eq.\,(\ref{eq:evol60plus}).}    
    
  \citet{SNStats} discussed that there seem to be two ways for LBVs,
either by exploding or turning into another WR stage to eventually explode as SN\,Ib
or Ic. Since the results of our WC analyses suggest that WNE and WC stars 
have progenitors of lower mass, this decision conflict would be solved. 
For an initial mass range from approximately 20\,$M_{\odot}$ to 50\,$M_{\odot}$, 
the WR stars would arise as post-RSG stars with most of their hydrogen being 
lost during the RSG stage, possibly skipping a hydrogen-rich WN stage that 
is not observed for this mass range. The evolutionary sequence leading to 
WC stars might therefore be
  
  \begin{equation}
    \label{eq:evol20to50}
    \mathrm{O} \rightarrow \mathrm{RSG} \rightarrow \mathrm{WNE} \rightarrow \mathrm{WC} \rightarrow \mathrm{SN\,Ib/c}.
  \end{equation}
  
  However, there is an upper mass limit for red supergiants. Although the 
exact value is unclear, it might be somewhere between 25\,$M_{\odot}$ and
50\,$M_{\odot}$ \citep[e.g.][]{Geneva2003,SmarttReview}, meaning that stars 
with higher initial masses do not become RSGs. Stars in the range above the
RSG mass limit but below the WNL mass range might therefore undergo an LBV
phase, but do not appear as WNL stars at any earlier stage. In addition, 
these ``lower mass LBVs" do not seem to explode as type II supernovae but
instead lose their hydrogen layers and eventually become hydrogen-poor
stars, such as WNE, WC, and WO stars. Assuming a mass limit for red
supergiants significantly below 50\,$M_{\odot}$, the evolutionary sequence
for the WC stars above would be

  \begin{equation}
    \label{eq:evol20to50lbv}
    \mathrm{O} \rightarrow \mathrm{LBV} \rightarrow \mathrm{WNE} \rightarrow \mathrm{WC} \rightarrow \mathrm{SN\,Ib/c}.
  \end{equation}
  
  Our empirical HRD positions for the Galactic WC stars instead support
the post-RSG origin of Eq.\,(\ref{eq:evol20to50}). All WC stars are located
below the 40\,$M_{\odot}$ tracks, independently of their detailed scenario 
assumptions. Even if the RSG upper mass limit is somewhat lower, most if not 
all WC stars of our sample must be attributed to the post-RSG evolution.

  Regardless of how the WC stars are produced, if they
explode as a SN, it clearly has to be of type Ib or Ic, as these stars are
hydrogen-free. While there are known progenitor stars for type II supernovae, 
there are none for type Ib and Ic. Therefore it is not yet clear if WNE or 
WC stars will undergo a supernova explosion at all. Instead they might just 
collapse to a black hole without a high luminosity outburst.
    
  To support the mass ranges for this new sequences, we take a closer
look at the discussions from \citet{SNStats}, assuming that WC stars 
always end up in a supernova. They tried to assign the different core 
collapse supernova types to mass ranges, so that the fractions fit with 
a Salpeter IMF. With the assumption all supernovae arise from single-star 
evolution they end up with a maximum mass of 22\,$M_{\odot}$
for type II supernovae, even including the SNe\,IIn here. That would
imply that all stars above 22\,$M_{\odot}$ would completely lose their
hydrogen, probably becoming a WNE and finally a WC or WO star. Although
this range would fit with the lower end of our WC initial mass range, this
scenario ignores the previously suggested WNL evolution and contradicts 
not only the idea, but also the observation of a SNe\,IIn with a high-mass 
progenitor.
  
  Even though they are only referring to core-collapse supernovae, it is 
clear that binaries have to be taken into account. In contrast to the
assumption that only single-star evolution dominates the SNe statistic,
one could assume that binary evolution is the only possible way for type Ib/c
SNe. That would of course assume that all WNE, WC, and WO stars are
binary systems or have at least been formed by close binary evolution.
Although a number of systems might have formed in this
way, it is highly unlikely that all WR stars (except WNL stars)
share this origin. Stellar evolution models might not cover the right mass 
range for these stars yet, but they do show that WNE and WC stars can be
formed without requiring close binary evolution.
  
  \citet{SNStats} therefore presented two ad hoc scenarios combining the
ideas of single star and binary evolution. A fraction of 8.8\% of their SN
sample are of type IIn while a total fraction of 26\% are type I CCSNe 
(hereafter labeled SNe\,Ib/c). Among these 26\%, some originate 
from single stars and others from binaries. As an ad hoc approach, 
they assumed that the single star fraction should be as large as
the SNe\,IIn fraction, namely 8.8\%. The remaining fraction is considered as a
result of close binary evolution, together with the SNe\,IIb, which are actually 
type Ib SNe with small amounts of hydrogen visible only in the early stage of 
the SN. For the type II-P and II-L SNe associated with the low initial masses,
they ended up with an upper limit to $M_{\mathrm{init}}$ of 23.1$\,M_{\odot}$. 
Above this initial mass, all single stars explode as either type Ib/c or IIn 
supernovae. Were a strict mass limit assumed to separate the SNe\,Ib/c and 
IIn scenarios, this would be $M_{\mathrm{init}} = 37\,M_{\odot}$ under the 
assumptions stated above. \citet{SNStats} assigned the SNe\,IIn to the lower 
mass range between 23\,$M_{\odot}$ and 37\,$M_{\odot}$, but according to the 
results of this work and the observation of \citet{LBVSNex}, it seems
likely that the converse is true. The IIn might then 
belong to stars with $M_{\mathrm{init}} > 37\,M_{\odot}$, possibly following 
the evolutionary scenario (\ref{eq:evol60plus}), and the SNe\,Ib/c come from
WC stars with initial masses between 23 and 37\,$M_{\odot}$, which is not
so far from our proposal based on the HRD positions relative to the
evolutionary tracks.
  
  The limits of the mass ranges are of course only rough estimations.
Neither our track comparisons, nor the ad hoc assumption of the single
star fraction in the SNe\,Ib/c fraction are based on detailed
calculations. The limits therefore can change in both directions, e.g.
depending on the values of the RSG mass-loss rates and the number of binaries
causing type Ib/c SNe. A larger single-star fraction in the ad hoc
assumption would decrease the minimum initial mass, while a lower RSG mass
loss would raise it. As an additional channel leading to SNe of type Ib/c,
 \citet{SNStats} suggested underluminous H-poor stars. We consider this idea
as speculative, because these objects have never been observed. Admittedly, 
if they formed by Roche-lobe overflow in close-binary systems, such stars 
might escape detection. However, on the basis of the SN statistics, the 
latter case is covered by the binary fraction that was already subtracted 
from the number of Ib/c supernovae.
 Despite all the remaining uncertainties, the scenario demonstrates 
that the observed CCSN statistics in general can be explained 
in-line with our empirical results.

  We have so far neglected two stars that do not seem to follow one of
the newly introduced scenarios (\ref{eq:evol60plus}) or
(\ref{eq:evol20to50}), namely the two WO2 stars WR 102 and WR 142. In
contrast to the early WC subtypes they appear as hot, but also very
luminous stars. Their positions are close to the end of the tracks of
\citet{Geneva2003} with favoring the non-rotational models. Their
positions therefore indicate that these two WO stars at least are stars
with $M_{\mathrm{init}} > 40\,M_{\odot}$. The end of each Geneva track
is marked by a SN symbol, which is not entirely appropriate as the
track calculations end when there is no more helium left in
the stellar core. It has been suggested that WO stars might
already be in a state of carbon burning \citep[e.g.][]{BH82}. The low
number of known WO stars -- just four of them in the Milky Way -- and 
the high temperature of the corresponding PoWR model compared to the
Geneva track endings support this idea. However, owing to the absence of
unambiguously identified helium lines in the optical spectrum, our models
might overestimate the helium abundance in the WO stars. Only if these 
stars were completely helium-free, we would have a clear evidence 
of a carbon-burning star.
  
  With the sequences (\ref{eq:evol60plus}) and (\ref{eq:evol20to50}), we
have evolutionary scenarios for the WC stars with masses of up to 
40\,$M_{\odot}$ or maybe 50\,$M_{\odot}$ and another for the WNL stars with
$M_{\mathrm{init}} > 60\,M_{\odot}$, suggested by the WNL positions and
the SN\,IIn constraints. The observed WC luminosities also 
roughly agree with the tracks of \citet{Vanbeveren98} for this mass range.
The tracks do not reproduce the obtained temperatures but this might be due 
to envelope inflation, which has never been included in any evolutionary
track to date. The WO stars might actually originate from stars with initial 
masses between the two mass bins. To end up as WO star, a star might fellow
a scenario close to the one of \citet{LangerSzenario}, which is 
also a kind of mixture of scenarios (\ref{eq:evol20to50}) and (\ref{eq:evol60plus})  
  \begin{equation}
    \label{eq:langerevol}
    \mathrm{O} \rightarrow \mathrm{WNL} \rightarrow \mathrm{LBV} \rightarrow \mathrm{WNE?} \rightarrow \mathrm{WO} \rightarrow \mathrm{SN\,Ib/c}.
  \end{equation}
In contrast to the original assumption of \citet{LangerSzenario}, the
WC phase is skipped here. If the WO stars were to evolve from WC stars, we
would expect to see WC stars with even higher luminosities than the WO
stars, which are not observed. The situation is not much better for WNE 
stars, hence this stage is tagged with a question mark 
in (\ref{eq:langerevol}) as there are no stars observed in a luminosity range
that would correspond to the luminosities of the WO stars. However, this might
be just a consequence of the low number statistics, as the WNE lifetimes are a
factor of between three and ten shorter than those of the WC stars, according 
to the calculations of \citet{Geneva2003}. In the case of He-burning
WO stars, their lifetimes would be comparable to those of WC stars and with 
only two known WO2 stars in the Milky Way, the absence of such WNE stars in the
corresponding luminosity range would be merely a statistical effect. 

  The final SN explosion, if there is one, should be of either type Ib or Ic. 
The latter would assume that such stars have lost all their helium. This is
not the case in our current WO models, which contain a helium mass fraction
of 30\%. However, as mentioned above, we cannot rule out that these stars 
might already be helium-depleted.

  Alternatively, one could imagine that WO stars form from rapidly rotating
WNL stars, with a very short or even without a WNE phase. This would be in 
line with our findings that both WO stars in our sample require a very high 
rotational velocity of 1000\,km/s to reproduce their round-shaped spectral 
lines. On the basis of these assumptions, the evolutionary 
scenario would be
  \begin{equation}
    \label{eq:worotevol}
    \mathrm{O} \rightarrow \mathrm{WNL} \rightarrow \mathrm{WO} \rightarrow \mathrm{GRB}.
  \end{equation}
Stellar evolution calculations currently support neither scenario (\ref{eq:langerevol})
nor (\ref{eq:worotevol}). The most recent Geneva models \citep{Geneva2011} 
predict a spin-down during the main-sequence stage, leading to lower 
rotational velocities at higher masses. However, we do see extremely broad 
lines in the WO spectra that can only explained by very rapid rotation. We 
speculate that these stars were either spun up by mass transfer from a 
companion, or that the transfer and loss of angular momentum remains 
inadequately described by the current evolutionary models.

  In total, the modified scenario of \citet{SNStats} corroborates our 
conclusion that WC stars arise from those of lower initial masses than
previously expected, as raised by the HRD positions. It is also in line
with the assumption that the very luminous WNL stars are core hydrogen-burning 
stars that will never reach the WNE and WC stages. However, the SN scenario
does not include metallicity effects and their binary assumptions are just
ad-hoc. As the formation of WR stars might strongly depend on the metallicity
of the host galaxy, the SN statistics show a plausible and interesting
scenario that is in line with our results, but one that should not be taken
as an independent argument for the WC mass range.

Metallicity effects may be illustrated by comparing WC stars in 
the Galaxy and the Large Magellanic Cloud (LMC). The six WC4 stars in the LMC
were analyzed by \citet{LMC1998} and \citet{CrowtherWC4LMC}.
According to the latter paper, the luminosities of these stars range from
$\log L = 5.4$ to $5.7$. \citet{LMC1998} obtained slightly lower values
(typically by 0.2\,dex), but used models without clumping and iron-line
blanketing at that time. Compared to the average luminosity of the WC4 stars
in our Galactic sample (see Table\,\ref{tab:wcmean}), it seems that their 
counterparts in the LMC are significantly brighter. The mass-loss rates, on 
the other hand,
are about the same (the assumptions on clumping are identical in both works),
which is obviously the reason why their line spectra are similar. One can 
conclude that (i) WC4 stars in the LMC have a higher current mass
than Galactic stars of the same type and (ii) a higher luminosity is needed at
LMC metallicity to drive the same mass loss. For probably related reasons,
WC stars of later subtypes and thus larger pseudo-photospheric radii do not
exist in the LMC.

\section{Conclusions}
  \label{sec:conclusions}

  We have analyzed 56 Galactic WC stars by comparing their
line spectra to simulations from a grid of line-blanketed PoWR model 
atmospheres to derive the stellar parameters. In
addition, the SED was fitted by adjusting the
reddening and scaling the luminosity to match flux-calibrated spectra
and photometric measurements. For the first time, a
clear correlation between the spectroscopic subtype and the stellar
parameters (Table\,\ref{tab:wcmean}) has been found. 

  Our results also show that for single WC stars the
subtypes are aligned along a strip in the $\log\,T_{*}$-$\log\,R_{\mathrm{t}}$-plane
with the WC9 stars having a slight offset. Moreover, in the $\dot{M}$-$L$-plane, 
we obtained an $\dot{M}$-$L$-relation for WC stars in the form of 
$\dot{M} \propto L^{0.8}$, which is in good agreement with \citet{NL2000mdot}. 

  For the chemical composition, the helium-to-carbon ratio seems to 
be rather uniform within the WC subclass (about 55:40 by mass). The oxygen
abundance is of the order of 5\%, but might be slightly higher for
the earliest subtypes and is about 30\% for the WO stars. 

  For the terminal wind velocity, we used values from \citet{Prinja90} and
\citet{NiSk2002} when available and scaled them for the remaining stars
before calculating the resulting stellar parameters.
Table\,\ref{tab:wcmean} reveals a significant
increase in the mean value of $\varv_{\infty}$ per subtype.
Starting with values of about $\varv_{\infty} \approx 1400$\,km/s for the
WC9 stars with some of them having even smaller values (close to 1000\,km/s),
$\varv_{\infty}$ quickly increases for earlier subtypes, ending
up as $\approx\,3300\,$km/s for WC4 stars. The masses obtained from
the luminosity via the relation of \citet{Langer89mass} seem to
roughly increase from late to early subtypes, but these values have to be
interpreted with care. The underlying mass-luminosity relation is indeed
independent of the scenario that led to the WR stage.

  The stars marked as WC9d display a significant excess even in the 
near-infrared, a finding that remains unexplained. Some of these 
stars might be part of a binary system. We obtained similar parameters 
as in our a binary pseudo fit of the well-known binary system WR 104
for some stars that are thought to be single stars. However, \cite{WC9cwb} 
found indications that not all WC9d-stars might have a companion.

\begin{table}
  \caption{Mean WC star parameters per subtype}
  \label{tab:wcmean}
  \centering

  \begin{tabular}{c r c c c c c }
      \hline
      \hline
\multicolumn{1}{c}{Subtype} & \multicolumn{1}{c}{$T_{*}$\rule[0mm]{0mm}{3mm}} & \multicolumn{1}{c}{$\varv_{\infty}$} & \multicolumn{1}{c}{$R_{*}$} & \multicolumn{1}{c}{$\log\,\dot{M}$\tablefootmark{a} } & \multicolumn{1}{c}{$\log\,L$} & \multicolumn{1}{c}{$M_{\mathrm{WC}}$\tablefootmark{b}} \\
& \multicolumn{1}{c}{[kK]} & \multicolumn{1}{c}{[km/s]} & \multicolumn{1}{c}{[$R_{\odot}$]} & \multicolumn{1}{c}{[$M_{\odot}$/yr]} & \multicolumn{1}{c}{[$L_{\odot}$]} & \multicolumn{1}{c}{[$M_{\odot}$]} \\
      \hline
WO2 \rule[0mm]{0mm}{3mm} & $200$ & $5000$ & $0.6$ & $-5.06$ & $5.7$ & $19$  \\
WC4 \rule[0mm]{0mm}{3mm} & $117$ & $3310$ & $1.0$ & $-4.65$ & $5.2$ & $10$  \\
WC5 \rule[0mm]{0mm}{3mm} & $83$ & $2780$ & $2.2$ & $-4.62$ & $5.3$ & $12$  \\
WC6 \rule[0mm]{0mm}{3mm} & $78$ & $2270$ & $2.9$ & $-4.61$ & $5.5$ & $14$  \\
WC7 \rule[0mm]{0mm}{3mm} & $71$ & $2010$ & $2.9$ & $-4.79$ & $5.3$ & $11$  \\
WC8 \rule[0mm]{0mm}{3mm} & $60$ & $1810$ & $4.2$ & $-4.80$ & $5.3$ & $11$  \\
WC9 \rule[0mm]{0mm}{3mm} & $44$ & $1390$ & $6.6$ & $-4.80$ & $5.2$ & $10$  \\[1mm]
     \hline
  \end{tabular}
  \tablefoot{
    \tablefoottext{a}{Mass-loss rates are calculated with a volume-filling factor of $f_{\mathrm{V}} = 0.1$}
    \tablefoottext{b}{WC masses are calculated from luminosities using the relation of \citet{Langer89mass}}
  }
\end{table}
  
  The WO stars WR\,102 and WR\,142 (both of WO2 subtype) differ 
significantly from the WC stars. They contain much more oxygen 
($X_{\mathrm{O}} \approx 30\%$) and have both much higher temperatures 
($T_{*} \approx 200\,$kK), and extremely fast winds with $\varv_{\infty} 
\approx 5000\,$km/s. They are very compact objects with masses higher 
than those of the WC stars and stellar radii smaller than 1\,$R_{\odot}$. 
Their round emission-line profiles indicate that these stars are rapid 
rotators. 

  The WN/WC stars are WN stars with an enhanced carbon fraction that 
have WN-type spectra. They are fitted with special models that
reflect these compositions, revealing that their basic parameters 
($T_{*}$, $R_{\mathrm{t}}$) are already close to those of the WC sequence. 
The star WR\,126 is sometimes thrown into this scheme because it is classified 
as WC/WN. In all cases, WR\,126 does not fit into any of the former groups, 
and seems to be an exception that has either its own unique history or is a 
so far undetected binary system, this last possibility being supported
by its proximity to the binary-pseudo-fit region in the 
$\log\,T_{*}$-$\log\,R_{\mathrm{t}}$-plane.
  
  The obtained positions of the WC stars in the HRD challenge 
the standard scenario that WC stars generally arise from very massive stars. 
Our results instead indicate that WC stars mostly originate from an 
intermediate-high mass range with initial masses of between $20\,M_{\odot}$
and $40\,M_{\odot}$ or maybe $50\,M_{\odot}$. This suggests that they are 
descendants of post-RSG or - for the higher masses - LBV stars, which have 
lost enough of their hydrogen envelope that they directly enter the WNE 
stage before they eventually become WC stars. 

  Although rotation decreases the minimum mass that is required to reach 
the WR stage, it seems that the scenarios without initial rotation provide
a closer fit to the observations. For the WC stars, the tracks of
\citet{Vanbeveren98} did not include rotation, but instead higher RSG 
mass-loss rates. These tracks are in much better agreement with our WC 
results than the Geneva tracks. The positions of the WO stars in the HRD are 
closer to the non-rotational Geneva tracks of \citet{Geneva2003}, which
appears to contradict the afore mentioned results for their line profiles, 
but it could simply indicate that fast rotation occurs only during the 
later stages of evolution.
  
  The evolutionary status of the WC9 stars is quite obscure, at least if they 
are single stars. They might originate from stars with initial masses of 
around 20\,$M_{\odot}$, but their evolutionary fate is completely unclear. 
As a first approach, one could imagine that they evolve to earlier WC 
subtypes, but that would have to be consistent with their increasing 
luminosity, as for early WC subtypes the luminosities are much higher than
for the WC9 stars. Such an increase in luminosity would contradict 
the current assumptions of the mass-luminosity relation that correlates
larger masses with larger luminosities.
  
  The WNL stars, using the term for WN stars containing hydrogen, seem to 
be different from the rest of the WR stars. Their HRD positions suggest 
that these stars might not actually be in the phase of helium burning 
traditionally expected of all WR stars. They might instead simply 
be core-hydrogen burning stars with very high masses and mass-loss rates, 
making them a kind of extension of the O stars at the high-mass end of
the main sequence. They are therefore sometimes not considered as 
``real" Wolf-Rayet stars. While this might seem logical in the way of using 
the term Wolf-Rayet stars for helium-burning stars, it is incorrect, 
as WR stars are not defined in terms of their core burning status but only 
their spectral characteristics. Nevertheless, the differences in the 
evolutionary status between the WNL and the other WR stars should be 
carefully taken into account for future stellar evolution calculations.

  The WO stars, at least those of WO2 subtype, seem to differ significantly
from WC stars, as we conclude from the different parameters and
HRD positions. However, their evolutionary status remains unclear. As
a tentative explanation, we suggest the scenarios (\ref{eq:langerevol}) 
and (\ref{eq:worotevol}). With only four known WO stars in our Galaxy, the 
observational basis is much worse than for either WN or WC stars, especially
as our results indicate that WO stars are not descendants of WC stars.
Scenario (\ref{eq:langerevol}) probably leads only in special situations
to a WO star, making WO stars such a rare subclass. 
  
When comparing the obtained HRD positions with evolutionary
models, we see a significant temperature discrepancy for WC5 stars and all 
later subtypes. A promising candidate to solve this problem is envelope
inflation, as described for H-free stars in \citet{Inflation2011}. 
Unlike the later subtypes, the WC4 stars and also the very different WO stars 
do not seem to be inflated.
  
\begin{table}
  \caption{Suggested single-star evolution scenarios based on WN and WC analyses}
  \label{tab:wcevol}
  \centering
  \begin{tabular}{r c l}
    \hline\hline
    \multicolumn{2}{c}{$M_{\mathrm{init}}$ [$M_{\odot}$] \rule[0mm]{0mm}{3mm}}  
                             &  \multicolumn{1}{c}{stellar evolution} \\ \hline
    \rule[0mm]{0mm}{3mm}     
      8~--~15	&	&		OB $\rightarrow$	RSG	$\rightarrow$		SN\,II-P															\\
     15~--~20	&	&		OB $\rightarrow$	RSG	$\rightarrow$		BSG				 $\rightarrow$	SN\,II-L		\\
     20~--~45	&	&		 O $\rightarrow$	RSG	$\rightarrow$		WNE				 $\rightarrow$	WC   $\rightarrow$ SN\,Ib/c	\\
     45~--~60	&	& 	 O $\rightarrow$	WNL	$\rightarrow$		LBV/WNE?	 $\rightarrow$	WO   $\rightarrow$ SN\,Ib/c	\\
       $>$~60	&	&	   O $\rightarrow$	Of/WNL	$\leftrightarrow$	LBV	[$\rightarrow$  WNL] $\rightarrow$ SN\,IIn 	\\
    \hline
  \end{tabular}
\end{table}

  A summary of the suggested single-star evolutionary scenarios for massive
stars based on the discussion in Sect.\,\ref{sec:evolution} can be found in
Table\,\ref{tab:wcevol}. As we found the upper mass limit for the WC stars 
to be somewhere in-between 40\,$M_{\odot}$ and 50\,$M_{\odot}$, we simply 
give the mean value of 45\,$M_{\odot}$. All of the mass limits specified in
Table\,\ref{tab:wcevol} are just rough suggestions, which are based on our
results and the referenced comparisons to SN statistics and stellar evolution
models. 

  Detailed stellar evolution models, including higher RSG mass-loss rates
and rotation, would enable us to check whether our suggested 
evolutionary scenarios can be numerically reproduced. More reliable distance
determinations for the Galactic WR stars would help to verify the low
luminosities obtained in this work. Detailed analyses of WC and WO stars in 
other galaxies would help us to circumvent the distance problems and enhance
our proposed scenario in order to check how it depends on metallicity.
  
  Both the WC and WO star parameters can provide constraints on the 
wind-driving mechanisms and the WC mass-loss rates can influence the chemical 
evolution of our Galaxy. Furthermore, WC and WO stars are keystones in the
present theories for the evolution of massive stars. In the past years, our 
picture of massive stars has changed and new evolutionary scenarios have been 
proposed. We hope that our results and suggestions can contribute to these 
discussions and help us to better understand the origin and evolution not only
of the WC stars, but of Wolf-Rayet stars in general.
  
  \textit{PoWR model remark:} The model grid used for the Galactic WC
stars will be published on our PoWR website\footnote{\texttt{http://www.astro.physik.uni-potsdam.de/PoWR.html}} 
with the model files being ready-to-use. The interface will allow you to 
select and view the basic model parameters before downloading the model itself.


\begin{acknowledgements}
  We acknowledge the helpful suggestions of the anonymous referee. We would 
like to thank Dany Vanbeveren for providing us with additional evolutionary
track material. This research has made use of the SIMBAD database, operated 
at CDS, Strasbourg, France. Most UV spectra used in this work are based on
INES data from the IUE satellite.   This publication makes use of data
products from the Two Micron All Sky Survey, which is a joint project of
the University of Massachusetts and the Infrared Processing and Analysis
Center/California Institute of Technology, funded by the National
Aeronautics and Space Administration and the National Science Foundation.
\end{acknowledgements}


\bibliographystyle{aa} 
\bibliography{wcpaper}

\begin{thebibliography}{114}
\expandafter\ifx\csname natexlab\endcsname\relax\def\natexlab#1{#1}\fi

\bibitem[{{Ardila} {et~al.}(2010){Ardila}, {van Dyk}, {Makowiecki}, {Stauffer},
  {Song}, {Rho}, {Fajardo-Acosta}, {Hoard}, \& {Wachter}}]{SASS}
{Ardila}, D.~R., {van Dyk}, S.~D., {Makowiecki}, W., {et~al.} 2010, VizieR
  Online Data Catalog, 219, 10301

\bibitem[{{Arnal}(1992)}]{Arnal92}
{Arnal}, E.~M. 1992, \aap, 254, 305

\bibitem[{{Bakker} \& {Th{\'e}}(1983)}]{BT1983}
{Bakker}, R. \& {Th{\'e}}, P.~S. 1983, \aaps, 52, 27

\bibitem[{{Barlow} \& {Hummer}(1982)}]{BH82}
{Barlow}, M.~J. \& {Hummer}, D.~G. 1982, in IAU Symposium, Vol.~99, Wolf-Rayet
  Stars: Observations, Physics, Evolution, ed. {C.~W.~H.~De Loore \&
  A.~J.~Willis} (Springer Netherlands), 387--392

\bibitem[{{Barniske} {et~al.}(2006){Barniske}, {Hamann}, \&
  {Gr{\"a}fener}}]{Barniske06}
{Barniske}, A., {Hamann}, W., \& {Gr{\"a}fener}, G. 2006, in Astronomical
  Society of the Pacific Conference Series, Vol. 353, Stellar Evolution at Low
  Metallicity: Mass Loss, Explosions, Cosmology, ed. {H.~J.~G.~L.~M.~Lamers,
  N.~Langer, T.~Nugis, \& K.~Annuk} (San Francisco: ASP), 243

\bibitem[{{Cardelli} {et~al.}(1989){Cardelli}, {Clayton}, \&
  {Mathis}}]{Cardelli}
{Cardelli}, J.~A., {Clayton}, G.~C., \& {Mathis}, J.~S. 1989, \apj, 345, 245

\bibitem[{{Chapman} {et~al.}(1999){Chapman}, {Leitherer}, {Koribalski},
  {Bouter}, \& {Storey}}]{Chapman99}
{Chapman}, J.~M., {Leitherer}, C., {Koribalski}, B., {Bouter}, R., \& {Storey},
  M. 1999, \apj, 518, 890

\bibitem[{{Cherepashchuk} \& {Karetnikov}(2003)}]{BinOrbits2003}
{Cherepashchuk}, A.~M. \& {Karetnikov}, V.~G. 2003, Astronomy Reports, 47, 38

\bibitem[{{Chu} \& {Treffers}(1981)}]{WR52dist}
{Chu}, Y. \& {Treffers}, R.~R. 1981, \apj, 250, 615

\bibitem[{{Conti}(1979)}]{Conti79}
{Conti}, P.~S. 1979, in IAU Symposium, Vol.~83, Mass Loss and Evolution of
  O-Type Stars, ed. {P.~S.~Conti \& C.~W.~H.~De Loore} (Springer Netherlands),
  431--443

\bibitem[{{Conti} \& {Vacca}(1990)}]{ContiVacca1990}
{Conti}, P.~S. \& {Vacca}, W.~D. 1990, \aj, 100, 431

\bibitem[{{Crowther} \& {Walborn}(2011)}]{OfWNClassf}
{Crowther}, P. \& {Walborn}, N. 2011, ArXiv e-prints

\bibitem[{{Crowther} {et~al.}(1998){Crowther}, {De Marco}, \&
  {Barlow}}]{WCClassCrowther}
{Crowther}, P.~A., {De Marco}, O., \& {Barlow}, M.~J. 1998, \mnras, 296, 367

\bibitem[{{Crowther} {et~al.}(2002){Crowther}, {Dessart}, {Hillier}, {Abbott},
  \& {Fullerton}}]{CrowtherWC4LMC}
{Crowther}, P.~A., {Dessart}, L., {Hillier}, D.~J., {Abbott}, J.~B., \&
  {Fullerton}, A.~W. 2002, \aap, 392, 653

\bibitem[{{Crowther} {et~al.}(2006){Crowther}, {Morris}, \&
  {Smith}}]{Crowther2006}
{Crowther}, P.~A., {Morris}, P.~W., \& {Smith}, J.~D. 2006, \apj, 636, 1033

\bibitem[{{de Jager} {et~al.}(1988){de Jager}, {Nieuwenhuijzen}, \& {van der
  Hucht}}]{deJager88}
{de Jager}, C., {Nieuwenhuijzen}, H., \& {van der Hucht}, K.~A. 1988, \aaps,
  72, 259

\bibitem[{{Dessart} {et~al.}(2000){Dessart}, {Crowther}, {Hillier}, {Willis},
  {Morris}, \& {van der Hucht}}]{Dessart2000}
{Dessart}, L., {Crowther}, P.~A., {Hillier}, D.~J., {et~al.} 2000, \mnras, 315,
  407

\bibitem[{{Dopita} {et~al.}(1990){Dopita}, {McGregor}, {Rawlings}, \&
  {Lozinskaia}}]{WR102dist}
{Dopita}, M.~A., {McGregor}, P.~J., {Rawlings}, S.~J., \& {Lozinskaia}, T.~A.
  1990, \apj, 351, 563

\bibitem[{{Dougherty} {et~al.}(2000){Dougherty}, {Williams}, \&
  {Pollacco}}]{DWP2000}
{Dougherty}, S.~M., {Williams}, P.~M., \& {Pollacco}, D.~L. 2000, \mnras, 316,
  143

\bibitem[{{Dougherty} {et~al.}(1996){Dougherty}, {Williams}, {van der Hucht},
  {Bode}, \& {Davis}}]{WR146cwb}
{Dougherty}, S.~M., {Williams}, P.~M., {van der Hucht}, K.~A., {Bode}, M.~F.,
  \& {Davis}, R.~J. 1996, \mnras, 280, 963

\bibitem[{{Drew} {et~al.}(2004){Drew}, {Barlow}, {Unruh}, {Parker}, {Wesson},
  {Pierce}, {Masheder}, \& {Phillipps}}]{WR93bfund}
{Drew}, J.~E., {Barlow}, M.~J., {Unruh}, Y.~C., {et~al.} 2004, \mnras, 351, 206

\bibitem[{{Dwarkadas}(2011)}]{Dwarkadas2011}
{Dwarkadas}, V.~V. 2011, \mnras, 412, 1639

\bibitem[{{Egan} {et~al.}(2003){Egan}, {Price}, {Kraemer}, {Mizuno}, {Carey},
  {Wright}, {Engelke}, {Cohen}, \& {Gugliotti}}]{MSX}
{Egan}, M.~P., {Price}, S.~D., {Kraemer}, K.~E., {et~al.} 2003, VizieR Online
  Data Catalog, 5114, 0

\bibitem[{{Fitzpatrick}(1999)}]{Fitzpatrick}
{Fitzpatrick}, E.~L. 1999, \pasp, 111, 63

\bibitem[{{Gal-Yam} \& {Leonard}(2009)}]{LBVSNex}
{Gal-Yam}, A. \& {Leonard}, D.~C. 2009, \nat, 458, 865

\bibitem[{{Gal-Yam} {et~al.}(2007){Gal-Yam}, {Leonard}, {Fox}, {Cenko},
  {Soderberg}, {Moon}, {Sand}, {Li}, {Filippenko}, {Aldering}, \&
  {Copin}}]{LBVSN}
{Gal-Yam}, A., {Leonard}, D.~C., {Fox}, D.~B., {et~al.} 2007, \apj, 656, 372

\bibitem[{{Gamen} \& {Niemela}(2003)}]{WR98bin}
{Gamen}, R.~C. \& {Niemela}, V.~S. 2003, in IAU Symposium, Vol. 212, A Massive
  Star Odyssey: From Main Sequence to Supernova, ed. {K.~van der Hucht,
  A.~Herrero, \& C.~Esteban} (San Francisco: ASP), 184

\bibitem[{{Garmany} \& {Stencel}(1992)}]{GS92}
{Garmany}, C.~D. \& {Stencel}, R.~E. 1992, \aaps, 94, 211

\bibitem[{{Gr{\"a}fener} \& {Hamann}(2005)}]{GH05}
{Gr{\"a}fener}, G. \& {Hamann}, W. 2005, \aap, 432, 633

\bibitem[{{Gr{\"a}fener} {et~al.}(1998){Gr{\"a}fener}, {Hamann}, {Hillier}, \&
  {Koesterke}}]{LMC1998}
{Gr{\"a}fener}, G., {Hamann}, W., {Hillier}, D.~J., \& {Koesterke}, L. 1998,
  \aap, 329, 190

\bibitem[{{Gr{\"a}fener} {et~al.}(2002){Gr{\"a}fener}, {Koesterke}, \&
  {Hamann}}]{GKH02}
{Gr{\"a}fener}, G., {Koesterke}, L., \& {Hamann}, W. 2002, \aap, 387, 244

\bibitem[{{Gr{\"a}fener} {et~al.}(2012){Gr{\"a}fener}, {Owocki}, \&
  {Vink}}]{Inflation2011}
{Gr{\"a}fener}, G., {Owocki}, S.~P., \& {Vink}, J.~S. 2012, \aap, 538, A40

\bibitem[{{Hamann} {et~al.}(2008){Hamann}, {Feldmeier}, \&
  {Oskinova}}]{ClumpingWS2008}
{Hamann}, W., {Feldmeier}, A., \& {Oskinova}, L.~M., eds. 2008, {Clumping in
  hot-star winds} (Universit{\"a}tsverlag Potsdam)

\bibitem[{{Hamann} \& {Gr{\"a}fener}(2004)}]{HG04}
{Hamann}, W. \& {Gr{\"a}fener}, G. 2004, \aap, 427, 697

\bibitem[{{Hamann} {et~al.}(2003){Hamann}, {Gr{\"a}fener}, \&
  {Koesterke}}]{HGK03}
{Hamann}, W., {Gr{\"a}fener}, G., \& {Koesterke}, L. 2003, in IAU Symposium,
  Vol. 212, A Massive Star Odyssey: From Main Sequence to Supernova, ed.
  {K.~van der Hucht, A.~Herrero, \& C.~Esteban} (San Francisco: ASP), 198

\bibitem[{{Hamann} {et~al.}(2006){Hamann}, {Gr{\"a}fener}, \&
  {Liermann}}]{Adriane06}
{Hamann}, W., {Gr{\"a}fener}, G., \& {Liermann}, A. 2006, \aap, 457, 1015

\bibitem[{{Hamann} \& {Koesterke}(1998)}]{HK98}
{Hamann}, W. \& {Koesterke}, L. 1998, \aap, 335, 1003

\bibitem[{{Hamann} {et~al.}(1995){Hamann}, {Koesterke}, \&
  {Wessolowski}}]{HKW95}
{Hamann}, W., {Koesterke}, L., \& {Wessolowski}, U. 1995, \aap, 299, 151

\bibitem[{{Hamann} {et~al.}(1992){Hamann}, {Leuenhagen}, {Koesterke}, \&
  {Wessolowski}}]{HLKW92}
{Hamann}, W., {Leuenhagen}, U., {Koesterke}, L., \& {Wessolowski}, U. 1992,
  \aap, 255, 200

\bibitem[{{Hamann} \& {Schmutz}(1987)}]{HS87}
{Hamann}, W. \& {Schmutz}, W. 1987, \aap, 174, 173

\bibitem[{{Hamann} {et~al.}(1993){Hamann}, {Koesterke}, \&
  {Wessolowski}}]{HKW93}
{Hamann}, W.~R., {Koesterke}, L., \& {Wessolowski}, U. 1993, \aap, 274, 397

\bibitem[{{Hillier}(1987)}]{Hillier87}
{Hillier}, D.~J. 1987, \apjs, 63, 947

\bibitem[{{Hillier}(1989)}]{Hi89}
{Hillier}, D.~J. 1989, \apj, 347, 392

\bibitem[{{Hillier} \& {Miller}(1998)}]{HiMi98}
{Hillier}, D.~J. \& {Miller}, D.~L. 1998, \apj, 496, 407

\bibitem[{{Hillier} \& {Miller}(1999)}]{HiMi99}
{Hillier}, D.~J. \& {Miller}, D.~L. 1999, \apj, 519, 354

\bibitem[{{Ishii} {et~al.}(1999){Ishii}, {Ueno}, \& {Kato}}]{Inflation1999}
{Ishii}, M., {Ueno}, M., \& {Kato}, M. 1999, \pasj, 51, 417

\bibitem[{{Kn{\"o}dlseder} {et~al.}(2002){Kn{\"o}dlseder}, {Cervi{\~n}o}, {Le
  Duigou}, {Meynet}, {Schaerer}, \& {von Ballmoos}}]{WR142olddistKn}
{Kn{\"o}dlseder}, J., {Cervi{\~n}o}, M., {Le Duigou}, J.-M., {et~al.} 2002,
  \aap, 390, 945

\bibitem[{{Koesterke} \& {Hamann}(1995)}]{KH95}
{Koesterke}, L. \& {Hamann}, W. 1995, \aap, 299, 503

\bibitem[{{Koesterke} {et~al.}(1992){Koesterke}, {Hamann}, \&
  {Wessolowski}}]{KH92}
{Koesterke}, L., {Hamann}, W., \& {Wessolowski}, U. 1992, \aap, 261, 535

\bibitem[{{Kotak} \& {Vink}(2006)}]{LBVSNSuggestion}
{Kotak}, R. \& {Vink}, J.~S. 2006, \aap, 460, L5

\bibitem[{{Langer}(1989)}]{Langer89mass}
{Langer}, N. 1989, \aap, 210, 93

\bibitem[{{Langer} {et~al.}(1994){Langer}, {Hamann}, {Lennon}, {Najarro},
  {Pauldrach}, \& {Puls}}]{LangerSzenario}
{Langer}, N., {Hamann}, W., {Lennon}, M., {et~al.} 1994, \aap, 290, 819

\bibitem[{{Liermann} {et~al.}(2010){Liermann}, {Hamann}, {Oskinova}, {Todt}, \&
  {Butler}}]{Adriane2010}
{Liermann}, A., {Hamann}, W.-R., {Oskinova}, L.~M., {Todt}, H., \& {Butler}, K.
  2010, \aap, 524, A82

\bibitem[{{Linder} {et~al.}(2009){Linder}, {Rauw}, {Manfroid}, {Damerdji}, {De
  Becker}, {Eenens}, {Royer}, \& {Vreux}}]{Linder09}
{Linder}, N., {Rauw}, G., {Manfroid}, J., {et~al.} 2009, \aap, 495, 231

\bibitem[{{Lortet} {et~al.}(1987){Lortet}, {Georgelin}, \&
  {Georgelin}}]{Lortet87}
{Lortet}, M.-C., {Georgelin}, Y.~P., \& {Georgelin}, Y.~M. 1987, \aap, 180, 65

\bibitem[{{Lundstr{\"o}m} \& {Stenholm}(1984)}]{LS84}
{Lundstr{\"o}m}, I. \& {Stenholm}, B. 1984, \aaps, 58, 163

\bibitem[{{Lynga}(1968)}]{Lynga1968}
{Lynga}, G. 1968, The Observatory, 88, 20

\bibitem[{{Maeder} \& {Meynet}(2011)}]{Geneva2011}
{Maeder}, A. \& {Meynet}, G. 2011, ArXiv e-prints

\bibitem[{{Mart{\'{\i}}n} {et~al.}(2007){Mart{\'{\i}}n}, {Cappa}, \&
  {Testori}}]{WR53Bubble}
{Mart{\'{\i}}n}, M.~C., {Cappa}, C.~E., \& {Testori}, J.~C. 2007, Revista
  Mexicana de Astronomia y Astrofisica, 43, 243

\bibitem[{{Martins} {et~al.}(2008){Martins}, {Hillier}, {Paumard},
  {Eisenhauer}, {Ott}, \& {Genzel}}]{Martins2008}
{Martins}, F., {Hillier}, D.~J., {Paumard}, T., {et~al.} 2008, \aap, 478, 219

\bibitem[{{Massey} {et~al.}(2001){Massey}, {DeGioia-Eastwood}, \&
  {Waterhouse}}]{WR142olddistMa}
{Massey}, P., {DeGioia-Eastwood}, K., \& {Waterhouse}, E. 2001, \aj, 121, 1050

\bibitem[{{Massey} \& {Johnson}(1993)}]{MJ93}
{Massey}, P. \& {Johnson}, J. 1993, \aj, 105, 980

\bibitem[{{Meynet} {et~al.}(2011){Meynet}, {Georgy}, {Hirschi}, {Maeder},
  {Massey}, {Przybilla}, \& {Nieva}}]{Meynet2011}
{Meynet}, G., {Georgy}, C., {Hirschi}, R., {et~al.} 2011, Bulletin de la
  Societe Royale des Sciences de Liege, 80, 266

\bibitem[{{Meynet} \& {Maeder}(2003)}]{Geneva2003}
{Meynet}, G. \& {Maeder}, A. 2003, \aap, 404, 975

\bibitem[{{Miller} \& {Chu}(1993)}]{WR145neb}
{Miller}, G.~J. \& {Chu}, Y. 1993, \apjs, 85, 137

\bibitem[{{Moffat} {et~al.}(1977){Moffat}, {Fitzgerald}, \&
  {Jackson}}]{Moffat1977}
{Moffat}, A.~F.~J., {Fitzgerald}, M.~P., \& {Jackson}, P.~D. 1977, \apj, 215,
  106

\bibitem[{{Moffat} \& {Vogt}(1973)}]{MoffatVogt1973}
{Moffat}, A.~F.~J. \& {Vogt}, N. 1973, \aap, 23, 317

\bibitem[{{Naz{\'e}} {et~al.}(2008){Naz{\'e}}, {Rauw}, \&
  {Manfroid}}]{WR20aPaper}
{Naz{\'e}}, Y., {Rauw}, G., \& {Manfroid}, J. 2008, \aap, 483, 171

\bibitem[{{Niedzielski} \& {Skorzynski}(2002)}]{NiSk2002}
{Niedzielski}, A. \& {Skorzynski}, W. 2002, \actaa, 52, 81

\bibitem[{{Niemela}(1991)}]{Niemela91}
{Niemela}, V.~S. 1991, in IAU Symposium, Vol. 143, Wolf-Rayet Stars and
  Interrelations with Other Massive Stars in Galaxies, ed. {K.~A.~van der Hucht
  \& B.~Hidayat} (Springer Netherlands), 201

\bibitem[{{Nugis} \& {Lamers}(2000)}]{NL2000mdot}
{Nugis}, T. \& {Lamers}, H.~J.~G.~L.~M. 2000, \aap, 360, 227

\bibitem[{{Oskinova} \& {Hamann}(2008)}]{WR65cwb}
{Oskinova}, L.~M. \& {Hamann}, W. 2008, \mnras, 390, L78

\bibitem[{{Oskinova} {et~al.}(2009){Oskinova}, {Hamann}, {Feldmeier}, {Ignace},
  \& {Chu}}]{WR142Lida}
{Oskinova}, L.~M., {Hamann}, W., {Feldmeier}, A., {Ignace}, R., \& {Chu}, Y.
  2009, \apjl, 693, L44

\bibitem[{{Pasemann} {et~al.}(2011){Pasemann}, {R{\"u}hling}, \&
  {Hamann}}]{DianaSMC}
{Pasemann}, D., {R{\"u}hling}, U., \& {Hamann}, W. 2011, Bulletin de la Societe
  Royale des Sciences de Liege, 80, 180

\bibitem[{{Petrovic} {et~al.}(2006){Petrovic}, {Pols}, \&
  {Langer}}]{Inflation2006}
{Petrovic}, J., {Pols}, O., \& {Langer}, N. 2006, \aap, 450, 219

\bibitem[{{Prinja} {et~al.}(1990){Prinja}, {Barlow}, \& {Howarth}}]{Prinja90}
{Prinja}, R.~K., {Barlow}, M.~J., \& {Howarth}, I.~D. 1990, \apj, 361, 607

\bibitem[{{Przybilla} {et~al.}(2010){Przybilla}, {Firnstein}, {Nieva},
  {Meynet}, \& {Maeder}}]{Przybilla2010}
{Przybilla}, N., {Firnstein}, M., {Nieva}, M.~F., {Meynet}, G., \& {Maeder}, A.
  2010, \aap, 517, A38

\bibitem[{{Radoslavova}(1989)}]{WR126dist}
{Radoslavova}, T. 1989, Astronomische Nachrichten, 310, 223

\bibitem[{{R{\"u}hling}(2008)}]{Ute08}
{R{\"u}hling}, U. 2008, Diplomarbeit, Universit{\"a}t Potsdam

\bibitem[{{Rustamov} \& {Cherepashchuk}(1989)}]{RC89}
{Rustamov}, D.~N. \& {Cherepashchuk}, A.~M. 1989, \azh, 66, 67

\bibitem[{{Schmutz} {et~al.}(1989){Schmutz}, {Hamann}, \&
  {Wessolowski}}]{SHW89}
{Schmutz}, W., {Hamann}, W., \& {Wessolowski}, U. 1989, \aap, 210, 236

\bibitem[{{Seaton}(1979)}]{Seaton}
{Seaton}, M.~J. 1979, \mnras, 187, 73P

\bibitem[{{Shorlin} {et~al.}(2004){Shorlin}, {Turner}, \&
  {Pedreros}}]{Shorlin2004}
{Shorlin}, S.~L., {Turner}, D.~G., \& {Pedreros}, M.~H. 2004, \pasp, 116, 170

\bibitem[{{Shylaja}(1990)}]{WR14bin}
{Shylaja}, B.~S. 1990, \apss, 164, 63

\bibitem[{{Skrutskie} {et~al.}(2006){Skrutskie}, {Cutri}, {Stiening},
  {Weinberg}, {Schneider}, {Carpenter}, {Beichman}, {Capps}, {Chester},
  {Elias}, {Huchra}, {Liebert}, {Lonsdale}, {Monet}, {Price}, {Seitzer},
  {Jarrett}, {Kirkpatrick}, {Gizis}, {Howard}, {Evans}, {Fowler}, {Fullmer},
  {Hurt}, {Light}, {Kopan}, {Marsh}, {McCallon}, {Tam}, {Van Dyk}, \&
  {Wheelock}}]{2MASS}
{Skrutskie}, M.~F., {Cutri}, R.~M., {Stiening}, R., {et~al.} 2006, \aj, 131,
  1163

\bibitem[{{Smartt}(2009)}]{SmarttReview}
{Smartt}, S.~J. 2009, \araa, 47, 63

\bibitem[{{Smartt} {et~al.}(2009){Smartt}, {Eldridge}, {Crockett}, \&
  {Maund}}]{Smartt2009}
{Smartt}, S.~J., {Eldridge}, J.~J., {Crockett}, R.~M., \& {Maund}, J.~R. 2009,
  \mnras, 395, 1409

\bibitem[{{Smith}(1968)}]{Smith1968}
{Smith}, L.~F. 1968, \mnras, 140, 409

\bibitem[{{Smith} {et~al.}(1990){Smith}, {Shara}, \& {Moffat}}]{Smith1990}
{Smith}, L.~F., {Shara}, M.~M., \& {Moffat}, A.~F.~J. 1990, \apj, 358, 229

\bibitem[{{Smith}(2006)}]{CarinaDist}
{Smith}, N. 2006, \apj, 644, 1151

\bibitem[{{Smith}(2008)}]{SNSmith2008}
{Smith}, N. 2008, in IAU Symposium, Vol. 250, Massive stars as cosmic engines,
  ed. {F.~Bresolin, P.~A.~Crowther, \& J.~Puls} (Cambridge University Press),
  193--200

\bibitem[{{Smith} \& {Conti}(2008)}]{WNLLBV}
{Smith}, N. \& {Conti}, P.~S. 2008, \apj, 679, 1467

\bibitem[{{Smith} {et~al.}(2011){Smith}, {Li}, {Filippenko}, \&
  {Chornock}}]{SNStats}
{Smith}, N., {Li}, W., {Filippenko}, A.~V., \& {Chornock}, R. 2011, \mnras,
  412, 1522

\bibitem[{{Th{\'e}} \& {Stokes}(1970)}]{TS1970}
{Th{\'e}}, P.~S. \& {Stokes}, N. 1970, \aap, 5, 298

\bibitem[{{Torres} \& {Massey}(1987)}]{SpecTM87}
{Torres}, A.~V. \& {Massey}, P. 1987, \apjs, 65, 459

\bibitem[{{Trundle} {et~al.}(2008){Trundle}, {Kotak}, {Vink}, \&
  {Meikle}}]{LBVSNother}
{Trundle}, C., {Kotak}, R., {Vink}, J.~S., \& {Meikle}, W.~P.~S. 2008, \aap,
  483, L47

\bibitem[{{Turner}(1996)}]{Turner96}
{Turner}, D.~G. 1996, \aj, 111, 828

\bibitem[{{Turner} \& {Forbes}(1982)}]{WR142olddistT82}
{Turner}, D.~G. \& {Forbes}, D. 1982, \pasp, 94, 789

\bibitem[{{Turner} {et~al.}(2006){Turner}, {Rohanizadegan}, {Berdnikov}, \&
  {Pastukhova}}]{WR142dist}
{Turner}, D.~G., {Rohanizadegan}, M., {Berdnikov}, L.~N., \& {Pastukhova},
  E.~N. 2006, \pasp, 118, 1533

\bibitem[{{van der Hucht}(2001)}]{vdH2001}
{van der Hucht}, K.~A. 2001, New Astronomy Review, 45, 135

\bibitem[{{van der Hucht}(2006)}]{vdH2006}
{van der Hucht}, K.~A. 2006, \aap, 458, 453

\bibitem[{{van Genderen} {et~al.}(1991){van Genderen}, {Verheijen}, {van der
  Hucht}, {de Loore}, {Schwarz}, {van Esch}, {Greidanus}, {van der Heiden},
  {van Kampen}, {Kuulkers}, {Le Poole}, {Reijns}, {Robijn}, \&
  {Spijkstra}}]{vanGenderen1991}
{van Genderen}, A.~M., {Verheijen}, M.~A.~W., {van der Hucht}, K.~A., {et~al.}
  1991, in IAU Symposium, Vol. 143, Wolf-Rayet Stars and Interrelations with
  Other Massive Stars in Galaxies, ed. {K.~A.~van der Hucht \& B.~Hidayat}
  (Springer Netherlands), 129

\bibitem[{{van Loon} {et~al.}(2005){van Loon}, {Cioni}, {Zijlstra}, \&
  {Loup}}]{vanLoon2005}
{van Loon}, J.~T., {Cioni}, M.-R.~L., {Zijlstra}, A.~A., \& {Loup}, C. 2005,
  \aap, 438, 273

\bibitem[{{Vanbeveren} {et~al.}(1998){Vanbeveren}, {De Donder}, {van Bever},
  {van Rensbergen}, \& {De Loore}}]{Vanbeveren98}
{Vanbeveren}, D., {De Donder}, E., {van Bever}, J., {van Rensbergen}, W., \&
  {De Loore}, C. 1998, New Astronomy, 3, 443

\bibitem[{{Vanbeveren} {et~al.}(2007){Vanbeveren}, {Van Bever}, \&
  {Belkus}}]{Vanbeveren2007}
{Vanbeveren}, D., {Van Bever}, J., \& {Belkus}, H. 2007, \apjl, 662, L107

\bibitem[{{Varricatt} \& {Ashok}(2006)}]{WR143Binary}
{Varricatt}, W.~P. \& {Ashok}, N.~M. 2006, \mnras, 365, 127

\bibitem[{{V{\'a}zquez} {et~al.}(2005){V{\'a}zquez}, {Baume}, {Feinstein},
  {Nu{\~n}ez}, \& {Vergne}}]{WR50cluster}
{V{\'a}zquez}, R.~A., {Baume}, G.~L., {Feinstein}, C., {Nu{\~n}ez}, J.~A., \&
  {Vergne}, M.~M. 2005, \aap, 430, 471

\bibitem[{{Vazquez} {et~al.}(1995){Vazquez}, {Will}, {Prado}, \&
  {Feinstein}}]{Vazquez1995}
{Vazquez}, R.~A., {Will}, J.-M., {Prado}, P., \& {Feinstein}, A. 1995, \aaps,
  111, 85

\bibitem[{{Wallace} {et~al.}(2005){Wallace}, {Gies}, {Moffat}, {Shara}, \&
  {Niemela}}]{WR38Cluster}
{Wallace}, D.~J., {Gies}, D.~R., {Moffat}, A.~F.~J., {Shara}, M.~M., \&
  {Niemela}, V.~S. 2005, \aj, 130, 126

\bibitem[{{Wallace} {et~al.}(2002){Wallace}, {Moffat}, \& {Shara}}]{WR104pic}
{Wallace}, D.~J., {Moffat}, A.~F.~J., \& {Shara}, M.~M. 2002, in Astronomical
  Society of the Pacific Conference Series, Vol. 260, Interacting Winds from
  Massive Stars, ed. {A.~F.~J.~Moffat \& N.~St-Louis} (San Francisco: ASP), 407

\bibitem[{{Williams} {et~al.}(2001){Williams}, {Kidger}, {van der Hucht},
  {Morris}, {Tapia}, {Perinotto}, {Morbidelli}, {Fitzsimmons}, {Anthony},
  {Caldwell}, {Alonso}, \& {Wild}}]{WR137dist}
{Williams}, P.~M., {Kidger}, M.~R., {van der Hucht}, K.~A., {et~al.} 2001,
  \mnras, 324, 156

\bibitem[{{Williams} {et~al.}(1994){Williams}, {van der Hucht}, {Kidger},
  {Geballe}, \& {Bouchet}}]{WR125dust}
{Williams}, P.~M., {van der Hucht}, K.~A., {Kidger}, M.~R., {Geballe}, T.~R.,
  \& {Bouchet}, P. 1994, \mnras, 266, 247

\bibitem[{{Williams} {et~al.}(2005){Williams}, {van der Hucht}, \&
  {Rauw}}]{WC9cwb}
{Williams}, P.~M., {van der Hucht}, K.~A., \& {Rauw}, G. 2005, in Massive Stars
  and High-Energy Emission in OB Associations, ed. {G.~Rauw, Y.~Naz{\'e}, \&
  R.~Blomme Gosset, E.}, 65--68

\bibitem[{{Williams} {et~al.}(1987){Williams}, {van der Hucht}, \&
  {The}}]{WvdHT1987}
{Williams}, P.~M., {van der Hucht}, K.~A., \& {The}, P.~S. 1987, \aap, 182, 91

\end{thebibliography}


\Online
\label{onlinematerial}

\begin{appendix}
  
\section{Comments on individual stars}
  \label{appsec:starcomments}

In this appendix we compile some additional information about
individual stars of our sample, especially their distance and binary status:

\textbf{WR\,4} is a WC5 star that is listed as SB1 in \citet{vdH2001}
owing to its short-periodic photometric variability measured by \citet{RC89},
but it displays neither diluted emission nor absorption lines
from a possible companion.  \citet{Smith1990} discussed whether WR\,4 belongs 
to an \ion{H}{i} bubble, for which they estimate a distance of between
1.6\,kpc and 2.73\,kpc depending on the method. The smaller value is 
close to our spectroscopic distance of 1.64\,kpc, based on the 
$M_{\varv}$ calibration.

\textbf{WR\,8} is classified as WN7/WCE. Its spectrum resembles
a WN7 type star, but with stronger carbon lines than usual. \citet{Niemela91}
measured large amplitude radial-velocity variations with a possible period
of 38.4\,d, where the carbon and nitrogen lines are in antiphase. This is 
indicative of a WN+WC binary system. We attempted to reproduce the spectrum
using a single-star model, which resembles a hydrogen-free WN star with enhanced
 carbon, but failed to obtain a convincing fit (Fig.\,B.3). 
\citet{LS84} estimated the modulus to be D.M. = 12.7\,mag based on
a membership of the Anon Pup\,a association.

\textbf{WR\,14} is a WC7 star, for which photometric variability has
been reported by various authors, and discussed by \citet{WR14bin} in
terms of a possible compact companion. It displays non-thermal radio
emission \citep{Chapman99}, which can normally be attributed to colliding 
winds in binary systems. We found that its spectrum can be reproduced with 
a typical WC model for single stars, and conclude that there is at least
no bright companion that dilutes the WC emission lines. \citet{LS84}
identified the star as a member of Anon Vel\,a, estimating a distance of
D.M. = 11.5\,mag.

\textbf{WR\,15} has the spectral type WC6. From the width of its lines,
we estimated a terminal wind velocity of 2600\,km\,s$^{-1}$, slightly
higher values than typical for this subtype. The distance of D.M. = 12.0\,mag is
based on the possible membership of Anon Vel\,b suggested by
\citet{LS84}. We prefer to used this distance method over that used to calculate
the spectroscopic parallaxes obtained by \citet{ContiVacca1990}, who derived 
a value of D.M. = 11.28\,mag for WR\,15.

\textbf{WR\,23} is another a WC6 star and belongs to the Car OB1 association,
for which \citet{CarinaDist} obtained a distance modulus of 
D.M. = 11.8\,mag for the Homunculus nebula around $\eta\,$Car. 
Earlier distance calculations for Car OB1 were D.M. = 12.55\,mag 
\citep{MJ93} and 12.1\,mag \citep{LS84}.

\textbf{WR\,26} is classified as WN7/WCE. The spectrum contains a very
strong \ion{C}{iv}\,5808\,\AA\ line, which requires a model with about a
20\% carbon mass fraction to be reproduced. The fit to our single-star
model is satisfactory. The high carbon abundance might indicate that 
WR\,26 is undergoing a transition from a WN to a WC star.

\textbf{WR\,33} has the spectral type WC5. The UV spectrum is indicative of 
a high wind velocity of about 3000\,km\,s$^{-1}$, but otherwise looks  
normal. The optical spectrum \citep[from][]{SpecTM87} shows very 
broad emission features between 4000\,\AA\ and 4600\,\AA\ that cannot be 
reproduced by any of our models.

\textbf{WR\,38} displays a WC4 spectrum. Our fit is not very consistent,
possibly indicating that the lines are diluted by a companion's
continuum. From a weak cluster around WR\,38, \citet{Shorlin2004}
derived a very large distance of D.M. = 15.8\,mag, which would imply a
very high luminosity of this star. In contrast, \citet{WR38Cluster}
obtained a distance modulus of $15.0^{+1.5}_{-1.0}$\,mag from the HST
photometry of the same cluster. Our photometric distance, based on the
subtype calibration, yields D.M. = 15.1\,mag, in perfect agreement 
with the latter work.

\textbf{WR\,39}, classified as WC7+OB? \citep{vdH2001}, is an interesting 
test case for the ``diluted emission-line" binarity criterion. When we analyze 
its line spectrum assuming that the star is single, the obtained parameters
place the star distinctly above the sequence of single WC stars in the
$\log\,T_{*}$-$\log R_{\mathrm{t}}$-plane (see Fig.\,\ref{fig:wcstats}).
Moreover, WR\,39 shows non-thermal radio emission \citep{Chapman99}, which is
normally attributed to colliding winds in binary systems. We therefore
conclude that WR\,39 is indeed most likely a binary.

\textbf{WR\,45} is a WC6 star. Unfortunately, the only spectrum 
available to us is limited to 3410--4730\,\AA\ and does not cover many
diagnostic lines. Therefore, the parameters we obtained for 
this star are uncertain.

\textbf{WR\,50} is a double-lined spectroscopic binary WC7+OB
\citep{vdH2001} with a light-curve period of 1d \citep{vanGenderen1991}. 
We analyzed its spectrum as if it were from a single star, and obtained 
parameters that are obviously affected by the d.e.l. effect (cf.\ 
Fig.\,\ref{fig:wcstats}). The distance modulus is D.M. = 12.8\,mag 
according to \citet{WR50cluster}.

\textbf{WR\,52} is one of only five known Galactic WC4 stars, and
does not show indications of binarity. The spectral fit
Fig.\,B.16) requires models with higher oxygen abundances
(10--15\% by mass) than we used for our standard WC grid (5\%). This
agrees with our findings of higher oxygen abundances for WC4 and
WO stars. As discussed by \citet{WR52dist}, a kinematic
distance of 2\,kpc and a photometric distance of 4\,kpc were
derived from the associated nebula. From our subtype calibration, we
obtained a distance modulus of D.M. = 12.67\,mag, corresponding to
3.4\,kpc.

\textbf{WR\,53} is classified as WC8 with the ``d" indicating
persistent dust. In contrast to the WC9 stars, dust formation 
is uncommon for this subtype, and might indicate colliding winds. 
The fit to our models, calculated with
the full \ion{C}{ii} ion, remains remarkably poor (Fig.\,B.17)
for unknown reasons. \citet{WR53Bubble} found an expanding \ion{H}{i} shell 
probably associated with this star, encircling an optical emission nebula, 
and estimated the distance of the shell to be 4$\pm$1\,kpc.
From our subtype calibration, we obtained 2.9\,kpc.

\textbf{WR\,58} is a transition-type star, classified as WN4/WC. For
this star, a carbon mass fraction of 0.1\% is enough to reproduce the 
\ion{C}{iv}\,5808\,\AA-line.

\textbf{WR\,59} is classified as WC9d. The star might be a binary, as
\citet{WC9cwb} found Balmer absorption features by
comparing the spectral lines to those of a non-dusty WC9 star.

\textbf{WR\,64} is classified as WC7. The only spectrum available to us 
is that of \citet{SpecTM87} and covers only a short wavelength range, 
making our analysis less precise. 

\textbf{WR\,65}, classified as WC9d, shows variable X-ray emission.
According to \citet{WR65cwb}, the emission originates from the
wind-wind collision in a massive binary system, and the variability
is caused by the different absorption columns along the orbit. 
\citet{WC9cwb} detected absorption features from the Balmer lines of 
hydrogen. We analyzed the spectrum as if it were from a single WC star 
and found no indication of a composite nature. However,
if WR\,65 were a member of the Cir\,OB1 association as suggested by 
\citet{Lortet87}, it would be by far the brightest WC star in our sample
($M_{\varv} = -7.02\,$mag adopting a distance modulus from \citet{Turner96} 
of 12.57\,mag). As this seems unlikely, we conclude that WR\,65 is located
in the foreground of Cir\,OB1, or it has a companion that contributes
significantly to its total brightness. In any case, we cannot employ
this star for our $M_{\varv}$ versus subtype calibration.

\textbf{WR\,68} is of subtype WC7 and a possible member of 
the Cir\,OB1 association. {\citet{Turner96} deduced a distance of 
D.M. = 12.57\,mag, based on the cluster Pismis 20, which is part of
Cir\,OB1.} Adopting this number leads to a plausible luminosity and HRD 
position. {However, the distance of Cir\,OB1 has been 
debated in the past with distance moduli ranging from 11.58\,mag
\citep{MoffatVogt1973} to 12.8\,mag \citep{Vazquez1995},
13.00\,mag \citep{Lortet87}, and 13.22\,mag \citep{Lynga1968}.}

\textbf{WR\,69} is one of the few WC9d stars for which a useful IUE
spectrum exists, since the reddening for this star is relatively
moderate. \citet{WC9cwb} measured differences in the radial velocities
between different observations, possibly indicating binarity.

\textbf{WR\,81} is another WC9 star, but one of the few Galactic
examples without dust emission. Our line fit is remarkably much more 
consistent than for the dusty WC9d stars.

\textbf{WR\,86} is a visual binary classified as WC7 that has a B0 
companion. The star is another example that helps to demonstrate which 
model parameters are obtained when a composite spectrum is fitted with
a single-star model. We again find that the ``diluted emission-line"
effect places the results from this pseudo-fit in the characteristic
domain of the $\log\,T_{*}$-$\log R_{\mathrm{t}}$-plane
(Fig.\,\ref{fig:wcstats}).

\textbf{WR\,88} is classified as WC9 and does not show dust emission.
Thus it resembles WR\,81. \citet{WC9cwb} attribute a couple of emission
lines to nitrogen, and conclude that WR\,88 is either a WC+WN binary or
belongs to a previously unobserved transitional WN/WC9 subtype. We
fitted the spectrum with our WC models, and obtained parameters that
agree with those of other WC9 stars (Fig.\,\ref{fig:wcstats}). Given the 
poor quality of the \citet{SpecTM87} spectrum we use, we refrain from test
calculations with enhanced nitrogen abundances.

\textbf{WR\,90} shows non-thermal radio emission \citep{Chapman99}, but
otherwise no indications of binarity. Our single-star fit and the
derived parameters are typical of its WC7 type.

\textbf{WR\,95} has the spectral subtype WC9d. According to
\citet{Moffat1977}, it belongs to the open cluster Trumpler\,27 and has a
distance of $2.1 \pm 0.2\,$kpc implying a distance modulus of 
D.M. = 11.61\,mag. \citet{TS1970} and \citet{BT1983} obtained lower 
values  of 10.17\,mag and 11.09\,mag, which would lead to very low luminosities.

\textbf{WR\,98} was originally classified as a single star of WN8/WC7
transition type. \citet{WR98bin} detected O8-9 type absorption lines
superimposed on its spectrum. From the SB2 radial velocity curves, they
derived the orbital elements. We tried to fit the spectrum with our
single-star models. According to the WN/WC transition type of the
spectrum, the best-fitting model is basically a hydrogen-free WN model
with enhanced carbon (5\% by mass). However, some spectral features
cannot be reproduced by any of our single-star models, most likely
because the contribution of the O-star companion cannot be neglected.
We therefore consider Fig.\,B.34 as a ``pseudo fit'' that
is in fact unsuitable for this composite spectrum, and omit WR\,98 from
our single-star analysis.

\textbf{WR\,102}, also known as Sand\,4, is one of two WO stars
analyzed in this work. Classified as WO2, it has the same subtype as WR\,142. 
To reproduce its spectrum, we needed to use models with an enhanced oxygen
abundance and very high wind velocities. Moreover, the
observed emission line profiles of WR\,142 have a round top, which differs
from the Gaussian-like shape usually encountered in WR spectra.
Our models can only reproduce this round shape when we convolve the
synthetic spectrum for rotational broadening with a $\varv \sin i$ of about
$1000$\,km\,s$^{-1}$. (For wind spectra, flux convolution is only a rough
approximation to account for rotation.) We note that the WO stars are very
compact, since their radius is smaller than the solar radius.
Interestingly, the same effect has also been found for the most compact
WN star in our Galaxy, the WN2 star WR\,2 \citep{Adriane06}. This
rapidly rotating, bare, and compact stellar core is certainly a 
gamma-ray burst candidate. For the distance of WR\,102, we adopt $3 \pm
1\,$kpc from \citet{WR102dist}, which is based on the nebula G2.4+1.4
that was originally classified as a supernova remnant, but is now
considered to be stellar ejecta from WR\,102 despite this star not being
located in its center. \citet{WR93bfund} calculated a distance 
of $4.6$\,kpc based on an IR photometry scaling relative to WR\,142.
We did not use this distance as it does not rely on an independent 
measurement. Using a value of $4.6$\,kpc would infer an extremely high 
luminosity of $\log L/L_{\odot}\,=\,6.1$.

\textbf{WR\,104} is well known for its pinwheel nebula. It is a binary 
system consisting of a WC9d and a B0.5V star. The latter is visually brighter
than the WC star \citep{WvdHT1987}. A third, fainter component was
resolved with HST \citep{WR104pic}, hence the WC emission lines are
expected to be ``diluted''. Nevertheless, an acceptable fit can still be
achieved for WR\,104 with a single-star model that has quite ``typical''
WC9 parameters (cf.\,Fig.\,\ref{fig:wcstats}). The distance (D.M. =
11.0\,mag) is adopted from \citet{LS84}, who assigned WR\,104 as a
possible member to the association Sgr\,OB1.

\textbf{WR\,111} is a prototypical WC5 star, which has been frequently
studied. \citet{GKH02} basically used the same models as the
present study, and therefore obtained similar results. \citet{GH05}
constructed a hydrodynamically consistent model for WR\,111, thus
showed for the first time that WC winds can be explained in terms of
radiation-driven mass loss. Their model provided a more consistent fit
of the line spectrum than the semi-empirical models used in the present
paper. The hydrodynamical model of \citet{GH05} has a much higher
stellar temperature $T_{\ast}$ = 140\,kK than our present study (89\,kK).
This mainly reflects the different radial structures of the two
respective models in the deepest zones of the wind -- we recall that
$T_\ast$ is defined as the effective temperature corresponding to the radius
$R_\ast$, where the Rosseland optical depth reaches 20
(cf.\,Eq.\ref{eq:lrt}). 

{The model of \citet{GH05} has nearly the same luminosity as
we obtain from our empirical fit. The mass-loss rate of the hydrodynamically
consistent model ($\log \dot{M} = -5.14$) is considerably lower than that of
our empirical model (-4.67). This difference is mainly due to the much higher
clumping contrast assumed by \citet{GH05} ($D$ = 50 in the
outer parts instead of 10 in this work). Such strong clumping is probably
unrealistic, but was needed to compensate for the incomplete line
opacities and achieve sufficient radiative driving. A minor part
of the mass-loss rate differences is due to the terminal wind velocity,
which was slightly smaller in the hydrodynamic model (2050\,km/s) than the value
adopted in the present paper (2398\,km/s). For the model of \citet{GH05},
a much higher stellar temperature ($T_{\ast}$ = 140\,kK) was assumed
than the 89\,kK given in the present paper. Different stellar radii $R_{\ast}$
compensate for the effect of luminosity. The similarity of the emergent
spectra demonstrate again the parameter degeneracy for very dense winds
discussed in Sect.\,\ref{subsec:param}}. The star 
WR\,111 is assumed to be a member of the Sgr\,OB1 association
\citep{LS84} with a distance modulus of D.M. = 11.0\,mag.

\textbf{WR\,113} is a WC8d+O8-9 binary system \citep{BinOrbits2003}, which
has an excess in its 2MASS K-band magnitude most probably caused by dust 
emission. The pseudo-fit of its spectrum obtained from
our single-star models leads to parameters in the binary domain of 
the $\log\,T_{*}$-$\log R_{\mathrm{t}}$-plane (Fig.\,\ref{fig:wcstats}), 
obviously due to the dilution of the emission lines. 

\textbf{WR\,114}, classified as WC5, is a member of the Ser\,OB1
association \citep{LS84} with a distance modulus of D.M. =
11.5\,mag. The spectrum was found to have diluted emission 
lines, and the star was therefore listed as a binary candidate in 
\citet{vdH2001}. However, our spectral fit and the obtained 
parameters are normal for a single WC5 star. 

\textbf{WR\,117} is a ``dusty'' WC9d star.  \citet{WC9cwb} did not find
any evidence of an OB companion, and our single-star model also fits
most features in the observed spectrum with typical parameters. The
lines are significantly stronger than in other WC9 stars. The
best-fitting model has a stellar temperature of $T_{*} = 56\,$kK, which
is relatively high for the WC9 subclass. The position of WR\,117 in the
$\log\,T_{*}$-$\log R_{\mathrm{t}}$-plane, as well as the high terminal
wind velocity, are close to or maybe already in the WC8 parameter
region. Some previous papers \citep[e.g.][]{ContiVacca1990} have indeed
classified WR\,117 as type WC8. Unfortunately, our available spectrum 
does not comprise any \ion{C}{ii}-lines, which would provide the criterion 
to distinguish the subtype WC8 from WC9.

\textbf{WR\,121} is another ``dusty'' WC9d star that is apparently
single. \citet{WC9cwb} could not find any evidence of an OB companion,
and our single-star model fits most features in the observed spectrum
with typical parameters. 

\textbf{WR\,125} is an SB2 binary system \citep[WC7ed+O9III,][]{WR125dust}, 
where the letters ``ed'' stand for episodic dust formation. Our pseudo
fit with a single-star model gives parameters that are characteristic of
composite spectra (cf.\ Fig.\,\ref{fig:wcstats}).

\textbf{WR\,126} shows a unique spectrum that differs from those of all 
other Wolf-Rayet subtypes. Its designation as WC5/WN indicates that it
has predominantly a WC-type spectrum (albeit the emission lines are
unusually weak), but also relatively strong nitrogen. We tentatively fit
the spectrum with a WC-type model from our low-carbon (20\% mass
fraction) grid (Fig.\,B.46). The \ion{N}{iv} lines at
7005--7031\,\AA\ reveal obviously a significant abundance of nitrogen,
which is not included in our WC models. The mass-loss rate is much lower than
those typically found for WC5 stars. A thorough spectral analysis of this
transition-type star is beyond the scope of the present paper. 
Following \citet{WR126dist}, the star is probably a member of the
Vul\,OB2 association, which has a distance modulus of D.M. = 13.2\,mag.

\textbf{WR\,132} is classified as WC6 and might be associated with an
\ion{H}{i}-bubble. Based on that bubble, \citet{Arnal92} estimated a
kinematic distance of 4.3\,kpc, which we adopt for our luminosity 
scaling.  

\textbf{WR\,135} is one of only four WC8 stars analyzed in this work.
The star is a possible member of Cyg\,OB3, leading to
a distance of D.M. = 11.2\,mag {\citep{GS92}. The earlier calculations
of \citet{LS84} obtained a slightly higher value of 11.6\,mag.} WR\,135 
is the only WC8 single star with an independent distance estimate.

\textbf{WR\,137} is a binary system (WC8pd+O9) showing periodic dust
(``pd'') formation. Under the assumption that the O star is a main-sequence 
star, \citet{WR137dist} estimated a distance modulus of D.M. =
11.1\,mag. The pseudo-fit with a single-star model yields parameters
that are typical for composite spectra (cf.\ Fig.\,\ref{fig:wcstats}).

\textbf{WR\,142}, classified as WO2, is the other WO star analyzed in
this work. It has the same subtype as WR\,102. The spectra of both stars
are very similar, hence so are the results of their analyses. We achieved the
best fit with a very high terminal wind velocity ($\varv_{\infty}$ =
5000\,km\,s$^{-1}$) and a very high stellar temperature of $T_{*}$ =
200\,kK. As for WR\,102, we had to convolve the emergent
spectrum with a high rotational broadening velocity of 1000\,km\,s$^{-1}$ 
in order to reproduce the round shape of the emission lines.
\citet{WR142Lida} detected weak but hard X-rays from this object. These
X-rays cannot be attributed to colliding stellar winds since there is
apparently no companion. WR\,142 is a member of the Berkeley 87
cluster. {The distance of this cluster has been disputed
in the past. \citet{WR142olddistT82} derived a cluster distance of
946$\pm$26\,pc, which would imply $\log\,L/L_{\odot} = 5.5$.
\citet{WR142olddistMa} and \citet{WR142olddistKn} obtained $1.58$\,kpc and
$1.8$\,kpc, raising the luminosity to 5.9 or even 6.0 for the latter
distance. In our work, we used the latest value of \citet{WR142dist}  
of a distance of 1230$\pm$40\,pc leading to $\log\,L/L_{\odot} = 5.7$.}

\textbf{WR\,143} was revealed as a binary (WC4+Be) by
\citet{WR143Binary}. The analysis with our single-star models yields 
parameters that closely fit to the WC sequence, possibly 
indicating that the contribution of the Be-type companion to the 
composite spectrum is relatively weak. 

\textbf{WR\,144} is one of two Galactic WC4 stars that are not suspected
to be binaries. According to \citet{LS84}, the star is a possible member of
the Cyg\,OB2 association with a distance modulus of 11.3\,mag, which we
adopt for our luminosity scaling. This distance implies a luminosity of
$\log L/L_{\odot} = 5.22$. An alternative distance modulus of
D.M. = 9.8\,mag for Cyg\,OB2 claimed by \citet{Linder09} would lead to an
implausibly low luminosity {of $\log L/L_{\odot} = 4.6$}.

\textbf{WR\,145} is of the transition type WN7/WC. Its spectrum is nicely
fitted by a WN-type model with enhanced carbon. The distance modulus of
11.3\,mag is taken from \citet{LS84} based on the possible membership
of WR\,145 in the Cyg OB2 association. The star is located in a nebula
\citep{WR145neb}.

\textbf{WR\,146} is a visual binary system consisting of a WC5 star with 
an O8 companion. \citet{WR146cwb} also resolved two components with a 
high-spatial-resolution radio observation and classified WR\,146 as a
colliding wind binary due to its non-thermal radio emission. \citet{DWP2000} 
performed an in-depth analysis of the different types of radio emission
from both components and their colliding wind region. They suggested that 
the companion might actually not be a single O8 star, but a composite system 
itself, possibly consisting of an O8 and another WC star to explain the high 
mass-loss rate. The poor pseudo fit that can be achieved with 
single-star models yields parameters which are typical of binaries if 
the d.e.l.\ effect is neglected (cf.\ Fig.\,\ref{fig:wcstats}). 
\citet{Dessart2000} tried to account for the companion's continuum when 
they analyzed this WR star, but still arrived at parameters that are 
atypical of a WC5 star. WR\,146 is also listed as a possible member 
of the Cyg\,OB2 association with D.M. = 11.3\,mag \citep{LS84}, although
this membership has since been questioned. \citet{WR146cwb} obtained D.M. 
= 10.4\,mag, which would place WR\,146 in front of Cyg\,OB2 and is the
value that we used in our pseudo fit.

\textbf{WR\,154} is a WC6 single star and a possible member of the 
Cep\,OB1 association, for which \citet{GS92} give a distance 
modulus of 12.2\,mag. {Earlier calculations obtained
D.M. = 13.7\,mag \citep{Smith1990} and 12.53\,mag \citep{ContiVacca1990}.}

\end{appendix}

\end{document}